\documentclass[aps,prc,twoside,twocolumn,nofootinbib,10pt,floatfix]{revtex4}
\usepackage{amsmath,amssymb}
\usepackage{graphicx,bm}
\usepackage{slashed}
\usepackage{epstopdf}
\usepackage{ulem} 
\usepackage[usenames]{color}
\usepackage{float}
\usepackage{hyperref}
\usepackage{subfigure}
\usepackage{rotating}
\usepackage{color}
\usepackage{multirow}
\usepackage{dcolumn}
\usepackage{overpic}
\usepackage{booktabs}
\usepackage{makecell}
\usepackage{arydshln} 
\usepackage{diagbox}
\usepackage{tabularx}
\usepackage{array}
\newcolumntype{|}{!{\vline}}

\renewcommand\sout{\bgroup \color{red} \ULdepth=-.5ex \ULset}

\newsavebox{\tablebox}
\usepackage{adjustbox}
\begin{document}
\title{Systematic investigation of the spectroscopy and decay behaviors of doubly-charmed pentaquarks
}
\author{Hong-Tao An$^{1}$}\email{anht@mail.tsinghua.edu.cn}
\author{Yu-Shuai Li$^{2}$}\email{liysh@pku.edu.cn}
\affiliation{
$^1$Department of Physics and Center for High Energy Physics, Tsinghua University, Beijing 100084, China \\
$^2$ School of Physics and Center of High Energy Physics, Peking University, Beijing 100871, China
}
\date{\today}
\begin{abstract}
Building upon the discoveries of the $\Xi^{++}_{cc}(3621)$ and $T^{+}_{cc}(3875)$,
we undertake a comprehensive investigation into the mass spectra, internal structures, and decay properties of doubly-charmed pentaquarks.
By treating the two light quarks as a tightly bound diquark, 
the five-body system reduces to a four-body heavy quark–heavy quark–diquark–antiquark configuration.
Within the constituent quark model framework, we calculate their mass spectra in the range of 4.7–5.4 GeV and corresponding internal mass contributions via the Gaussian expansion method.
The root-mean-square radii, typically between 1.1-1.6 fm, indicate compact spatial structures. 
Furthermore, we also calculate the rearrangement decay widths via the quark-interchange model,
finding that all states are unstable and decay into a singly-charmed baryon and a singly-charmed meson.
Several narrow resonances have been identified, some of which have a total width even below 10 MeV.
We hope that our study could provide valuable guidance for future theoretical investigations and experimental searches targeting doubly-charmed pentaquarks.
\end{abstract}
\maketitle

\section{Introduction}\label{sec1}
Since the observation of X(3872) \cite{Belle:2003nnu}, research focus in the field of hadron physics has increasingly shifted towards exotic states that exist beyond the conventional mesons ($q\bar{q}$) and baryons ($qqq$) configurations \cite{Chen:2016qju,Guo:2017jvc,Liu:2019zoy,Hosaka:2016pey,Chen:2022asf}.
Following the discovery of a series of exotic states such as XYZ states \cite{Belle:2011aa,LHCb:2022aki,CMS:2013jru,LHCb:2016axx,LHCb:2021uow,BESIII:2013ouc}, $P_{c(s)}$ states \cite{LHCb:2015yax,LHCb:2019kea,LHCb:2020jpq,LHCb:2022ogu}, and $T_{cc}$ state \cite{LHCb:2021vvq,LHCb:2022sfr}, these studies have emerged as a central topic of research interest, driven by the need to explore new forms of matter and deepen our understanding of strong interaction dynamics.
Numerous potential explanations have been proposed in the existing literature, with the most prominent including conventional hadrons in the unquenched picture \cite{Ferretti:2014xqa,Luo:2019qkm,Duan:2020tsx}, tetraquarks, pentaquarks (compact bound states) \cite{Yang:2020atz,Meng:2023jqk,Huang:2023jec},  
hadronic molecules (loosely bound states of several hadrons) \cite{Wang:2019ato,Dong:2021juy,Dong:2021bvy}, 
hybrids (composed of gluons and quarks) \cite{Brambilla:2019jfi,Farina:2020slb,Brambilla:2022hhi}, glueballs (composed solely of gluons, without quark-antiquark component) \cite{Crede:2008vw,Zhang:2022obn,Vereijken:2023jor}, and kinematic enhancement mechanisms \cite{Wang:2020kej,Wang:2020tpt,Braaten:2022elw}. 

In the field of doubly-charmed hadrons, 
the SELEX Collaboration first claimed the observation of the 
doubly-charmed baryon $\Xi^{+}_{cc}(3520)$ via the $pD^{+}K^{-}$ three-body decay channel in 2002 \cite{SELEX:2002wqn};
however, its existence has not been confirmed by other collaborations \cite{BaBar:2006bab,Belle:2006edu}. 
In 2017, the LHCb Collaboration discovered the 
doubly-charmed baryon  $\Xi^{++}_{cc}(3621)$ 
via the $\Lambda^{+}_{c}K^{-}\pi^{+}\pi^{+}$  channel \cite{LHCb:2017iph}.
Subsequently, the LHCb Collaboration has also observed several different decay modes of this baryon, such as $\Xi^{++}_{cc}\to\Xi^{(')+}_{c}\pi^{+}$ \cite{LHCb:2018pcs,LHCb:2022rpd}, $\Xi^{++}_{cc}\to\Xi^{0}_{c}\pi^{+}\pi^{+}$ \cite{LHCb:2025shu}...
Notably, other doubly heavy baryons have likewise been investigated by the LHCb Collaboration, yet no signals have been detected thus far \cite{LHCb:2021xba}.
In 2022, the LHCb Collaboration reported the discovery of the first doubly-charmed tetraquark candidate $T^{+}_{cc}(3875)$, in the $D^{0}D^{0}\pi^{+}$ mass distribution just below the $D^{*+}D^{0}$ threshold \cite{LHCb:2021vvq}.
The $T^{+}_{cc}(3875)$ is a doubly-charmed manifestly exotic state.
The simplest assumption on its valence quark component is $cc\bar{u}\bar{d}$, its width is $\Gamma=410$ keV and its quantum numbers were determined to be $I(J^{P})=0(1^{+})$ \cite{LHCb:2022sfr}.
The identification of $T^{+}_{cc}(3875)$ has laid a solid foundation for further exploration of the spectroscopy and dynamics of doubly-heavy hadronic systems. 

In fact, before the observation of the $T^{+}_{cc}(3875)$, 
theoretical studies for multiquark states with doubly heavy quarks
had been conducted for many years \cite{Du:2012wp,Chen:2013aba,Luo:2017eub}.
For instance, based on the discovered $\Xi^{++}_{cc}(3621)$,
Karliner {\it et al.} predicted a stable 
doubly-bottom tetraquark $T_{bb\bar{u}\bar{d}}$ with spin-parity $J^{P}=1^{+}$ \cite{Karliner:2017qjm,Eichten:2017ffp}.
This state is stable against the strong and electromagnetic (EM) interactions and can only weak decay.
After the observation of $T^{+}_{cc}(3875)$, 
the theoretical studies have been categorized into two main categories:  
one treats $T^{+}_{cc}(3875)$ as a compact tetraquark state composed of four valence quarks \cite{Kim:2022mpa,Abreu:2022lfy,Agaev:2021vur,Azizi:2021aib}, while the other interprets it as a $DD^{*}$ molecular state \cite{Chen:2021vhg,Meng:2021jnw,Du:2021zzh,He:2022rta}.

The discovery and theoretical investigations of $\Xi^{++}_{cc}(3621)$ and $T^{+}_{cc}(3875)$ have also spurred the many theoretical attempts to explore the existence and properties of doubly-charmed pentaquarks \cite{Zhou:2022gra,Duan:2024uuf,Chen:2025obf}.
For example, in Ref.\cite{Xing:2021yid}, Xing {\it et al.} studied the masses and lifetimes of doubly charmed pentaquarks in the doubly heavy triquark-diquark framework.
In Refs.\cite{Li:2025omw,Zhou:2018bkn}, Zhou {\it et al.} predicted several stable doubly-heavy pentaquarks within the frame of color-magnetic interaction model.
Additionally, Park {\it et al.} investigated doubly-charmed pentaquarks using a quark model with a complete set of harmonic oscillator bases.
Meanwhile, Yang {\it et al.} also investigated doubly heavy pentaquarks within the framework of the QCD sum rule method in Refs.\cite{Wang:2018lhz,Yang:2024okq}.
Beyond mass spectra and stability, Ref.\cite{Ozdem:2022vip} further focused on the electromagnetic properties of doubly heavy pentaquarks.
Furthermore, there are other discussions in the compact pentaquark configuration \cite{Andreev:2022qdu,Giannuzzi:2019esi,Park:2018oib,Yang:2020twg} and the hadronic molecule configuration \cite{Wang:2023mdj,Wang:2023aob,Chen:2021kad,Shen:2022zvd}.

When tackling many-body problems such as baryons or multiquark states, the diquark model not only simplifies the relevant calculations but has also been proven effective in hadron spectroscopy \cite{Anwar:2018sol,Anselmino:1992vg,Barabanov:2020jvn}. 
The concept of diquarks traces its origins to the foundational development of the quark model: initially introduced to provide an alternative description of baryons as bound states of a constituent quark and a diquark, it has since gained phenomenological support for the emergence of diquark-like correlations \cite{Lichtenberg:1969sxc,Lichtenberg:1967zz,Ida:1966ev}. 
Such evidence includes the Regge behaviour of hadrons, namely the fact that baryons and mesons can be accommodated on Regge trajectories with approximately the same slope \cite{Santopinto:2004hw,Johnson:1975sg}; 
$\Lambda(1116)$ and $\Lambda(1520)$ fragmentation functions \cite{Wilczek:2004im,Selem:2006nd}; 
and the absence of the $\Lambda(\frac{3}{2}^{+})$ baryon state from the baryon spectrum\cite{Jaffe:2004ph}... 
In particular, the light diquark-heavy quark picture has been widely applied to single heavy-flavor baryon systems ($cqq$) \cite{Ebert:2011kk,Chen:2016iyi,Chen:2017gnu,Chen:2018orb,Chen:2021eyk}, where the mass spectra can be accurately reproduced.
As a crucial extension of singly-charmed baryon spectroscopy,
doubly-charmed pentaquarks ($ccqq\bar{q}$) likewise comprise two light-flavor quarks.
Building on this structural similarity, 
we propose that their mass spectra also exhibit analogous symmetry when the same theoretical approach is employed. 
Accordingly, we further extend and apply the heavy quark-heavy quark-light diquark-antiquark configuration (a four-body framework) to the doubly charmed pentaquark system (a five-body system).
Specifically, by ``freezing" the internal degrees of freedom of the light diquark, this simplification reduces computational complexity while preserving critical physical features.

Inspired by the observed $\Xi^{++}_{cc}(3621)$ and $T^{+}_{cc}(3875)$ as well as the widespread application of the diquark model, 
we systematically investigate the doubly-charmed pentaquark system 
$ccqq\bar{q}$ ($q = n, s; n = u, d$) within the heavy quark-heavy quark-diquark-antiquark configuration by combining these two aspects.
The mass spectra of all possible flavor combinations are obtained via the Gaussian expansion method within the framework of the constituent quark model. 
Additionally, we also calculate the corresponding internal mass contributions, root-mean-square (RMS) radii, and rearrangement decay properties.

The paper is organized as follows. 
After the Introduction, Section \ref{sec2} details the theoretical framework, including the effective Hamiltonian, pentaquark configuration, Gaussian expansion method, root-mean-square (RMS) radii, and rearrangement decay properties. 
Then, the numerical results for mass spectra, internal structure, and decay properties, along with specific discussions are presented in Section \ref{sec3}
The paper concludes with a summary in Section \ref{sec4}.

\section{Theoretical framework for doubly-charmed pentaquark system}\label{sec2}
\subsection{The effective Hamiltonian
}\label{sec21}
To investigate the mass spectra, internal structure, and strong decay stability of multiquark systems, we adopt a nonrelativistic constituent quark model, which incorporates one-gluon-exchange potential. 
The general form of this Hamiltonian is given as follows \cite{Park:2016mez,Park:2017jbn,An:2022qpt,An:2022fvs}:
\begin{eqnarray}\label{Eq1}
H=\sum_{i=1}^{n}(m_{i}+\frac{\textbf{p}^{2}_{i}}{2m_{i}})-\frac{3}{4}\sum_{i<j}^{n}
\frac{\lambda^{c}_{i}}{2}.\frac{\lambda^{c}_{j}}{2}(V^{\rm C}_{ij}+V^{\rm CS}_{ij}),
\end{eqnarray}
where the confinement potential $V^{\rm C}$ and hyperfine potential $V^{\rm CS}$ are expressed as follows:
\begin{eqnarray}\label{Eq2}
V^{{\rm C}}_{ij}&=&-\frac{\kappa}{r_{ij}}+\frac{r_{ij}}{a^{2}_{0}}-D,\nonumber\\
V^{{\rm CS}}_{ij}&=&\frac{\kappa'}{m_{i}m_{j}}\frac{1}{r_{0ij}r_{ij}}e^{-r^{2}_{ij}/r^{2}_{0ij}}\sigma_{i}\cdot\sigma_{j}.
\end{eqnarray}
Here, $n$ is the number of quarks in the hadron and $m_{i}$ is the mass of $i$-th quark.
The $\textbf{p}^{2}_{i}/(2m_{i})$ stands for the kinetic energy of the $i$-th quark.
The $\lambda^{c}_{i}/2$ and $\sigma_{i}$ denote the $SU(3)$ color and $SU(2)$ spin operators, respectively.
For an antiquark, $\lambda^{c}_{i}$ should be replaced by $-\lambda^{c*}_{i}$. 
The $r_{ij}=|\textbf{r}_{i}-\textbf{r}_{j}|$ is the interquark distance between the $i$-th and the $j$-th quarks.
Regarding the parameters $r_{0ij}$ and $\kappa'$, we have
\begin{eqnarray}\label{Eq3}
r_{0ij}&=&1/(\alpha+\beta\frac{m_{i} m_{j}}{m_{i}+m_{j}}),\nonumber\\
\kappa'&=&\kappa_{0}(1+\gamma\frac{m_{i} m_{j}}{m_{i}+m_{j}}).
\end{eqnarray}
The numerical values of the parameters in Eqs. (\ref{Eq2}-\ref{Eq3}) are determined by fitting them to the experimental masses of the singly-charmed hadrons, and these parameters are listed in Table \ref{para}.
Here, we adopt the heavy quark- light diquark configuration to fit the theoretical masses of singly-charmed baryons, to ensure
configurational consistency for subsequent pentaquarks calculations.
For completeness, Table \ref{para} also includes the theoretical and experimental masses of singly-charmed hadrons, alongside their corresponding errors to facilitate comparative analysis.
\\

\begin{table*}[t]
\caption{Parameters of the Hamiltonian determined by fitting the singly-charmed hadron masses.
The $M_{\rm the}$, $M_{\rm exp}$, and Error are the theoretical value, the experimental value, and the error between them, respectively.
}\label{para}
\begin{lrbox}{\tablebox}
\renewcommand\arraystretch{1.85}
\renewcommand\tabcolsep{2.8pt}
\begin{tabular}{c|cccccccccc}
\toprule[1.50pt]
\toprule[0.50pt]
Parameter&\multicolumn{1}{c|}{$m_{n}$}&\multicolumn{1}{c|}{$m_{s}$}&\multicolumn{2}{c|}{$a_{0}$}&\multicolumn{2}{c|}{$\beta$}&
\multicolumn{2}{c}{$\kappa_{0}$}\\
\Xcline{1-9}{0.5pt}
Value&\multicolumn{1}{c|}{320.0 MeV}&\multicolumn{1}{c|}{647.0 MeV}&\multicolumn{2}{c|}{$2.0\times10^{-2}$ $\rm (MeV^{-1}fm)^{1/2}$}& \multicolumn{2}{c|}{$-5.7\times10^{-4}$ $\rm (MeV fm)^{-1}$}&\multicolumn{2}{c}{$2.9\times10^{2}$ MeV}\\
\Xcline{1-9}{0.5pt}
Parameter&\multicolumn{1}{c|}{$m_{c}$}&\multicolumn{1}{c|}{$D$}&\multicolumn{2}{c|}{$\alpha$}&\multicolumn{2}{c|}{$\kappa$}&\multicolumn{2}{c}{$\gamma$}\\
\Xcline{1-9}{0.5pt}
Value&\multicolumn{1}{c|}{1509.0 MeV}& \multicolumn{1}{c|}{1034.0 MeV}&\multicolumn{2}{c|}{1.3 $\rm fm^{-1}$}&\multicolumn{2}{c|}{$1.5\times10^{2}$ MeV fm}&\multicolumn{2}{c}{$-5.7\times10^{-4}$ $\rm MeV^{-1}$}\\
\midrule[1.5pt]
Baryon&\multicolumn{1}{c|}{$\Lambda_{c}$}&\multicolumn{1}{c|}{$\Sigma_{c}$}&\multicolumn{1}{c|}{$\Sigma^{*}_{c}$}&\multicolumn{1}{c|}{$\Xi_{c}$}&
\multicolumn{1}{c|}{$\Xi'_{c}$}&\multicolumn{1}{c|}{$\Xi^{*}_{c}$}&\multicolumn{1}{c|}{$\Omega_{c}$}&\multicolumn{1}{c}{$\Omega^{*}_{c}$}
\\ 
\midrule[0.5pt]
$M_{\rm the}$ (MeV)&\multicolumn{1}{c|}{2273.7}&\multicolumn{1}{c|}{2450.5}&
\multicolumn{1}{c|}{2530.7}&\multicolumn{1}{c|}{2487.0}&\multicolumn{1}{c|}{2577.9}&\multicolumn{1}{c|}{2647.3}&\multicolumn{1}{c|}{2685.4}&\multicolumn{1}{c}{2747.7}
\\ \midrule[0.5pt]
$M_{\rm exp}$ (MeV)&\multicolumn{1}{c|}{2286.5}&\multicolumn{1}{c|}{2452.9}&\multicolumn{1}{c|}{2517.5}&\multicolumn{1}{c|}{2467.8}&\multicolumn{1}{c|}{2577.4}&\multicolumn{1}{c|}{2645.9}&\multicolumn{1}{c|}{2695.2}& \multicolumn{1}{c}{2765.9}
\\ \midrule[0.5pt]
Error (MeV)&\multicolumn{1}{c|}{-12.8}&  \multicolumn{1}{c|}{-2.4}& \multicolumn{1}{c|}{\quad \quad 12.2\quad\quad\quad}&\multicolumn{1}{c|}{19.2}&\multicolumn{1}{c|}{\quad\quad 0.5 \quad\quad\quad}&\multicolumn{1}{c|}{ 1.4 }&\multicolumn{1}{c|}{ \quad-9.8\quad\quad}&\multicolumn{1}{c}{ -18.2 }\\
\midrule[1.0pt]
Meson&\multicolumn{1}{c|}{$D$}&\multicolumn{1}{c|}{$D^{*}$}&\multicolumn{1}{c|}{$D_{s}$}&\multicolumn{1}{c|}{$D^{*}_{s}$}\\
\Xcline{1-5}{0.5pt}
$M_{\rm the}$ (MeV)&\multicolumn{1}{c|}{1867.3}&\multicolumn{1}{c|}{2013.8}&\multicolumn{1}{c|}{1969.2}&\multicolumn{1}{c|}{2109.5}\\
\Xcline{1-5}{0.5pt}
$M_{\rm exp}$ (MeV)&\multicolumn{1}{c|}{1869.7}&\multicolumn{1}{c|}{2010.3}&\multicolumn{1}{c|}{1968.3}&\multicolumn{1}{c|}{2112.2}\\
\Xcline{1-5}{0.5pt}
Error (MeV)&\multicolumn{1}{c|}{-2.4}&\multicolumn{1}{c|}{3.5}&\multicolumn{1}{c|}{0.9}&\multicolumn{1}{c|}{-2.7}\\
\bottomrule[0.50pt]
\bottomrule[1.50pt]
\end{tabular}
\end{lrbox}\scalebox{1.05}{\usebox{\tablebox}}
\end{table*}

\subsection{Pentaquark configuration
}\label{sec22}

In order to calculate the mass spectra, root-mean-square radii, and rearrangement decay width of the doubly-charmed pentaquarks, 
we need to construct their total wave functions, which are the direct product of the spatial, flavor, color, and spin wave functions:
\begin{eqnarray}\label{total}
\Psi_{tot}=\Psi_{spatial}\otimes F_{flavor}\otimes \psi_{color}\otimes \chi_{spin}.
\end{eqnarray}

When two light quarks are treated as a tightly bound diquark [qq], 
the doubly-charmed pentaquarks reduce to a four-body system under the 
[$c$-$c$]-[diquark-antiquark] configuration.

In flavor space, six distinct flavor configurations exist for the doubly-charmed pentaquark system: $cc[nn]\bar{n}$, $cc[nn]\bar{s}$, $cc[ss]\bar{n}$, $cc[ss]\bar{s}$, $cc[ns]\bar{n}$, and $cc[ns]\bar{s}$.

In the color space, a diquark with the color-antitriplet configuration ($(qq)^{\bar{3}_{c}}$) is regarded as the ``good" diquark
because of its attractive confining potential. 
In contrast, the color-sextet diquark ($(qq)^{6_{c}}$), labeled the ``bad" diquark, displays repulsive interactions. 
Meanwhile, due to the requirement of color confinement, the color
wave function must be a singlet, and thus the color decomposition proceeds as:
\begin{eqnarray}\label{color}
&&(3\otimes3)\otimes([3\otimes3]\otimes\bar{3})\nonumber\\
&=&(3\otimes3)\otimes([\bar{3}\oplus6]\otimes\bar{3})\nonumber\\
&\to&(3\otimes3)\otimes(\bar{3}\otimes\bar{3})
=(\bar{3}\oplus6)\otimes(3\oplus\bar{6})\nonumber\\
&\to&(\bar{3}\otimes3)\oplus(6\otimes\bar{6}).
\end{eqnarray}
From Eq.~(\ref{color}), two color-singlet wave functions can be constructed:
\begin{eqnarray}\label{color1}
\psi_{1}=|(cc)^{\bar{3}_{c}}[(q_{1}q_{2})^{\bar{3}_{c}}\bar{q}]^{3_{c}}\rangle,
\psi_{2}=|(cc)^{6_{c}}[(q_{1}q_{2})^{\bar{3}_{c}}\bar{q}]^{\bar{6}_{c}}\rangle.
\end{eqnarray}
In the notation $|(cc)^{\rm color_{1}}[(q_{1}q_{2})^{\bar{3}_{c}}\bar{q}]^{\rm color_{2}}\rangle$,
the $\rm color_{1}$ and $\rm color_{2}$ stand for the color representations of charm-quark pair and $[q_{1}q_{2}]\bar{q}$ substructure, respectively.

In the spin space, the doubly-charmed pentaquark system supports 10 spin wave functions for the [$c$-$c$]-[diquark-antiquark] configuration.
All possible spin wave functions are tabulated in Table~\ref{spin}.
Here, in the notation $|(cc)_{\rm spin_{1}}[(q_{1}q_{2})_{\rm spin_{2}}\bar{q}_{\frac{1}{2}}]_{\rm spin_{3}}\rangle_{\rm spin_{4}}$,
$\rm spin_{1}$ and $\rm spin_{2}$ represent the spins of the charm-quark pair ($cc$) and diquarks ($q_{1}q_{2}$), respectively.
While, $\rm spin_{3}$ and $\rm spin_{4}$ are the total spin of $[q_{1}q_{2}]\bar{q}$ substructure and the whole doubly-charmed pentaquarks.

\begin{table}[t]
\centering
\caption{All possible spin wave functions of the doubly-charmed pentaquarks in the 
[$c$-$c$]-[diquark–antiquark] configuration.}\label{spin}
\begin{lrbox}{\tablebox}
\renewcommand\arraystretch{1.5}
\renewcommand\tabcolsep{2.5pt}
\begin{tabular}{c|cccccc}
\toprule[1.50pt]
\toprule[0.50pt]
\multicolumn{2}{l}{spin wave function:}\\
\toprule[0.50pt]
\multirow{1}*{$J=\frac{5}{2}$}
&\multicolumn{1}{r}{
$\chi_{1}=|(cc)_{s=1}[(nn)_{1}\bar{n}]_{s=\frac{3}{2}}\rangle_{s=\frac{5}{2}}$}\\
\midrule[0.5pt]
\multirow{2}*{$J=\frac{3}{2}$}
&\multicolumn{1}{r}{$\chi_{2}=|(cc)_{s=1}[(nn)_{1}\bar{n}]_{s=\frac{3}{2}}\rangle_{s=\frac{3}{2}}$}
&\multicolumn{1}{r}{$\chi_{3}=|(cc)_{s=1}[(nn)_{1}\bar{n}]_{s=\frac{1}{2}}\rangle_{s=\frac{3}{2}}$}
\\
&\multicolumn{1}{r}{$\chi_{4}=|(cc)_{s=0}[(nn)_{1}\bar{n}]_{s=\frac{3}{2}}\rangle_{s=\frac{3}{2}}$}
&\multicolumn{1}{r}{$\chi_{5}=|(cc)_{s=1}[(nn)_{0}\bar{n}]_{s=\frac{1}{2}}\rangle_{s=\frac{3}{2}}$}
\\
\midrule[0.5pt]
\multirow{3}*{$J=\frac{1}{2}$}
&\multicolumn{1}{r}{$\chi_{6}=|(cc)_{s=1}[(nn)_{1}\bar{n}]_{s=\frac{3}{2}}\rangle_{s=\frac{1}{2}}$}
&\multicolumn{1}{r}{$\chi_{7}=|(cc)_{s=1}[(nn)_{1}\bar{n}]_{s=\frac{1}{2}}\rangle_{s=\frac{1}{2}}$}
\\
&\multicolumn{1}{r}{$\chi_{8}=|(cc)_{s=0}[(nn)_{1}\bar{n}]_{s=\frac{1}{2}}\rangle_{s=\frac{1}{2}}$}
&\multicolumn{1}{r}{$\chi_{9}=|(cc)_{s=1}[(nn)_{0}\bar{n}]_{s=\frac{1}{2}}\rangle_{s=\frac{1}{2}}$}
\\
&\multicolumn{1}{r}{$\chi_{10}=|(cc)_{s=0}[(nn)_{0}\bar{n}]_{s=\frac{1}{2}}\rangle_{s=\frac{1}{2}}$}
\\
\bottomrule[0.50pt]
\bottomrule[1.50pt]
\end{tabular}
\end{lrbox}\scalebox{0.86}{\usebox{\tablebox}}
\end{table}

With the above preparation, we can begin to construct
$\Psi_{spatial}\otimes F_{flavor}\otimes\psi_{color}\otimes\chi_{spin}$ wave functions that satisfy the Pauli principle.
Here, we only consider low-lying $S$-wave pentaquarks, 
the spatial wave function is symmetric under the exchange of any two identical quarks.
Thus, the remaining components must be fully antisymmetric under the exchange of identical quarks.
Combining different flavor combinations, we present in Tables \ref{ccnnn}-\ref{ccnsn} all possible color-spin configurations with different $I(J^{P})$ quantum numbers that satisfy the Pauli principle.
Here, we use the notation $|(cc)^{\rm color_{1}}_{\rm spin_{1}}[(q_{1}q_{2})^{\bar{3}_{c}}_{\rm spin_{2}}\bar{q}_{\frac{1}{2}}]^{\rm color_{2}}_{\rm spin_{3}}\rangle_{\rm spin_{4}}$ to label the total wave function.

\subsection{Gaussian expansion method}\label{sec231}

To accurately solve the four-body problem, we employ the Gaussian expansion method (GEM) \cite{Hiyama:2003cu,Hiyama:2012sma}.
As a well-established variational approach, GEM has been successfully applied to quantum few-body problems in baryons \cite{Luo:2023sne,Luo:2023sra}, tetraquarks \cite{Yang:2025jsp,Wu:2021rrc}, pentaquarks \cite{Yan:2021glh,Yan:2023iie}, few-body molecular states \cite{Wu:2023eyd,Wu:2021kbu,Luo:2021ggs,Luo:2022cun}, as well as analogous systems in atomic and nuclear physics \cite{Naidon:2011vov,Hiyama:2012cj}.

In spatial space, the spatial wave function for a pentaquark system with zero angular momentum can be expanded using a set of correlated Gaussian basis functions.
The expression for the correlated Gaussian functions is given by:
\begin{eqnarray}\label{spatial}
\psi={\rm Exp}[-\sum^{4}_{i<j}a_{ij}(\vec{x}_{i}-\vec{x}_{j})^{2}],
\end{eqnarray}
where $a_{ij}$ are the variational parameters.
To decouple the center-of-mass and relative motion degrees of freedom, it is convenient to adopt a set of Jacobi coordinates 
$\xi=\{\xi_{1},\xi_{2},\xi_{3}\}$ instead of position vectors $X=\{\vec{x}_{1},\vec{x}_{2},\vec{x}_{3},\vec{x}_{4}\}$. 
Here, we use the set of Jacobi coordinates for a four-body system as follows (see Fig. \ref{fig2}):
\begin{eqnarray}\label{jacobi}
&&\xi_{1}=\sqrt{1/2}(\vec{x}_{1}-\vec{x}_{2}),\nonumber\\
&&\xi_{2}=\sqrt{1/2}(\vec{x}_{3}-\vec{x}_{4}),\nonumber\\
&&\xi_{3}=(\frac{m_{1}\vec{x}_{1}+m_{2}\vec{x}_{2}}{m_{1}+m_{2}})-(\frac{m_{3}\vec{x}_{3}+m_{4}\vec{x}_{4}}{m_{3}+m_{4}}),\\
&&\textbf{R}=\frac{m_{1}\vec{x}_{1}+m_{2}\vec{x}_{2}+m_{3}\vec{x}_{3}+m_{4}\vec{x}_{4}}{m_{1}+m_{2}+m_{3}+m_{4}}.\nonumber
\end{eqnarray}
Here, $\xi_{1}$ ($\xi_{2}$) stands for the relative Jacobi coordinate between the charm quarks $c$ (diquark $[q_{1}q_{2}]$ and antiquark $\bar{q}$).
Meanwhile, $\xi_{3}$ corresponds to the relative Jacobi coordinate between the centers of mass
of the two charm quarks $(cc)$ and $[q_{1}q_{2}]\bar{q}$ substructure.
Using the above Jacobi coordinates (as defined in Eq.~(\ref{jacobi})), the spatial wave function with well-defined symmetry properties for the $cc$ quark pair (12) can be readily constructed.

Moreover, in the center-of-mass frame of the four-body system ($\textbf{R}=0$),
the number of independent Jacobi coordinates is reduced to three.
The corresponding kinetic term of the Hamiltonian (Eq. (\ref{Eq1})) can be
simplified and expressed as:
\begin{eqnarray}\label{kinetic term}
T_{c}=\sum^{4}_{i=1}\frac{\textbf{p}^{2}_{\vec{x}_{i}}}{2m_{i}}- \frac{\textbf{p}^{2}_{R}}{2M}= \frac{\textbf{p}^{2}_{\xi_{1}}}{2m'_{1}}+\frac{\textbf{p}^{2}_{\xi_{2}}}{2m'_{2}}+\frac{\textbf{p}^{2}_{\xi_{3}}}{2m'_{3}},
\end{eqnarray}
where the reduced masses $m'_{i}$ (i=1,2,3) are defined as follows:
\begin{eqnarray}\label{kinetic term2}
m'_{1}&=&\frac{2\times (m_{1}m_{2})}{m_{1}+m_{2}}, \nonumber\\
m'_{2}&=&\frac{2\times (m_{3}m_{4})}{m_{3}+m_{4}}, \nonumber\\
m'_{3}&=&\frac{ (m_{1}+m_{2}) \times (m_{3}+m_{4})}{m_{1}+m_{2}+m_{3}+m_{4}}.
\end{eqnarray}
In addition, based on the above Jacobi coordinates (Eq. (\ref{jacobi})), the correlated Gaussian functions (Eq. (\ref{spatial})) can be rewritten as:
\begin{eqnarray}\label{spatial2}
\psi(\xi_{1},\xi_{2},\xi_{3})={\rm Exp}[-\sum_{i,j}A_{ij}\xi_{i}\cdot\xi_{j}]={\rm Exp}[-\tilde{\xi}A\xi],
\end{eqnarray}
where $A$ is a $3\times3$ symmetric positive-definite matrix and its matrix 
elements are the variational parameters.
Furthermore, the spatial part of the total wave function is constructed as a linear combination of the correlated Gaussian functions:
\begin{eqnarray}\label{spatial3}
&&\Psi_{spatial}(\xi_{1},\xi_{2},\xi_{3})\nonumber\\
&=&\sum^{n_{1max}}_{n_{1}=1}\sum^{n_{2max}}_{n_{2}=1}\sum^{n_{3max}}_{n_{3}=1}
c_{n_{1}n_{2}n_{3}}\psi^{n_{1}n_{2}n_{3}}(\xi_{1},\xi_{2},\xi_{3})\nonumber\\
&=&\sum^{n_{1max}}_{n_{1}=1}\sum^{n_{2max}}_{n_{2}=1}\sum^{n_{3max}}_{n_{3}=1}
c_{n_{1}n_{2}n_{3}}\rm Exp[-a_{n_{1}}\xi^{2}_{1}-a_{n_{2}}\xi^{2}_{2}-a_{n_{3}}\xi^{2}_{3}],\nonumber\\
\end{eqnarray}
where the expansion coefficients $c_{n_{1}n_{2}n_{3}}$ are determined via the Rayleigh-Ritz variational method.
The Gaussian range parameters $a_{n_{i}}$ (for i=1,2,3) are chosen via a geometric progression:
\begin{eqnarray}\label{spatial4}
a_{n_{i}}=\frac{1}{r^{2}_{n_{i}}}, \quad r_{n_{i}}=r_{min_{i}}d^{n_{i}-1}, 
\end{eqnarray}
where the ratio coefficient $d$ is given by
\begin{eqnarray}\label{spatial4}
d=(\frac{r_{max_{i}}}{r_{min_{i}}})^{\frac{1}{n_{max_{i}}-1}} \quad (i=1,2,3).
\end{eqnarray}
Here, $r_{max_{i}}$ and $r_{min_{i}}$ define the maximum and minimum spatial extents, while $n_{max_{i}}$ specifies the number of Gaussian basis functions.
These three parameters \{$r_{max_{i}}$, $r_{min_{i}}$, $n_{max_{i}}$\} are determined through the variation method to minimize the energy.
Stable numerical results are typically achieved with the choices \{5 fm, 0.7 fm, 5\}.
These results are independent of the parameters \{$r_{max_{i}}$, $r_{min_{i}}$, $n_{max_{i}}$\}.
To verify this parameter independence, we adjusted the value of $n_{max_{i}}$ from 4 to 6 and confirmed that the calculated energies remain consistent.

Following the previous preparations, 
the eigenvalues of doubly-charmed pentaquark system are determined by solving the four-body Schr\"{o}dinger equation:
\begin{eqnarray}\label{equa}
\hat{H}\Psi_{\rm tot}(\xi_{1},\xi_{2},\xi_{3})=E\Psi_{\rm tot}(\xi_{1},\xi_{2},\xi_{3}),
\end{eqnarray}
where $\hat{H}$ is the Hamiltonian operator and $E$ represents the energy eigenvalues. 
The $\hat{H}$ corresponds to Eq.~(\ref{Eq1}), consisting of the kinetic term (with the specific form given by Eq.~(\ref{kinetic term})) and the two-body interaction potentials between (di)quarks. Meanwhile, $\Psi_{\rm total}(\xi_{1},\xi_{2},\xi_{3})$ is the total wave function corresponding to Eq.~(\ref{total}), composed of the spatial part (with the specific form shown in Equation Eq.~(\ref{spatial3})) and the flavor-color-spin part $|(cc)^{\rm color_{1}}_{\rm spin_{1}}[(q_{1}q_{2})^{\bar{3}_{c}}_{\rm spin_{2}}\bar{q}_{\frac{1}{2}}]^{\rm color_{2}}_{\rm spin_{3}}\rangle_{\rm spin_{4}}$.

The matrix elements of kinetic, potential, and normalization are calculated as follows:
\begin{eqnarray}
\label{acca}
T_{c}^{nn'}&=&\langle\Psi^{n_{1}n_{2}n_{3}}(\xi_{1},\xi_{2},\xi_{3})\psi_{cs}|T_{c}|\psi'_{cs}\Psi^{n'_{1}n'_{2}n'_{3}}(\xi_{1},\xi_{2},\xi_{3})\rangle,\nonumber\\
V^{nn'}_\alpha&=&
\langle\Psi^{n_{1}n_{2}n_{3}}(\xi_{1},\xi_{2},\xi_{3})\psi_{cs}|
V_\alpha|\psi'_{cs}\Psi^{n'_{1}n'_{2}n'_{3}}(\xi_{1},\xi_{2},\xi_{3})\rangle ,\nonumber\\
N^{nn'}&=&\langle\Psi^{n_{1}n_{2}n_{3}}(\xi_{1},\xi_{2},\xi_{3})\psi_{cs}|\psi'_{cs}\Psi^{n'_{1}n'_{2}n'_{3}}(\xi_{1},\xi_{2},\xi_{3})\rangle.\nonumber\\
\end{eqnarray}
Here, $\psi_{cs}$ represents the spin-color wave function, $V$ corresponds to $V^{\rm Con}$ and $V^{\rm SS}$ as specified in Eq.~(\ref{Eq2}), 
and $n$  simply stands for $\{n_{1}, n_{2}, n_{3}\}$ given in Eq.~(\ref{spatial}). 
The $V_\alpha$ ($\alpha=$1-6) represents $V(x_{12})$, $V(x_{13})$, $V(x_{14})$, $V(x_{23})$, $V(x_{24})$, and $V(x_{34})$.
For more details of the derivation, refer to Ref.\cite{Brink:1998as}.

By virtue of Eq.~(\ref{acca}), Eq.~(\ref{equa}) can be converted into a generalized matrix eigenvalue problem,i.e.,
\begin{eqnarray} \label{tven}
[T^{nn'}_{c}+\sum^{6}_{\alpha=1}V^{nn'}_{\alpha}]C_{nn'}=EN^{nn'}C_{nn'}.
\end{eqnarray}
Using the above analytical expressions, 
we calculate the mass spectrum (the eigenvalue $E$) and the corresponding internal contributions for doubly-charmed pentaquarks, as summarized in Tables \ref{ccnnn}-\ref{ccnsn}.

\subsection{Root-mean-square radii}\label{sec232}

To gain deeper insights into the internal structure of the 
doubly-charmed pentaquark system, 
we further calculate the root-mean-square (RMS) radii for all (di)quark pairs.
Here, the RMS radius $\langle r^{2}_{ij} \rangle$ is defined as follows \cite{Liu:2024fnh}:
\begin{eqnarray}\label{rms}
\langle r^{2}_{ij} \rangle=\int (\textbf{x}_{i}-\textbf{x}_{j})^{2}|\Psi(\xi_{1},\xi_{2},\xi_{3})|^{2}d\xi_{1}d\xi_{2}d\xi_{3},
\end{eqnarray}
where $\langle r^{2}_{ij} \rangle^{\frac{1}{2}}$ represents the RMS spatial separation between the $i$-th (di)quark and $i$-th (anti)quark.
The corresponding results for different flavor configurations are also summarized in Tables \ref{ccnnn}-\ref{ccnsn}.
Specifically, the physical interpretation of each RMS radius is as follows:
$\langle r^{2}_{12} \rangle^{\frac{1}{2}}$ describes the RMS spatial separation between two charm quarks ($cc$);
$\langle r^{2}_{34} \rangle^{\frac{1}{2}}$ describes the RMS spatial separation between light diquark $[q_{1}q_{2}]$ and antiquark $\bar{q}$;
$\langle r^{2}_{13} \rangle^{\frac{1}{2}}$ and $\langle r^{2}_{23} \rangle^{\frac{1}{2}}$ describe the RMS spatial separation between charm quark $c$ and light diquark $[q_{1}q_{2}]$;
$\langle r^{2}_{14} \rangle^{\frac{1}{2}}$ and $\langle r^{2}_{24} \rangle^{\frac{1}{2}}$ describe the RMS spatial separation between charm quark $c$ and antiquark $\bar{q}$;
$\langle r^{2}_{12-34} \rangle^{\frac{1}{2}}$ describes the average distance between the mass centers of the charm quark pair ($cc$) and the $[q_{1}q_{2}]\bar{q}$ substructure; $\langle r^{2}_{13-24} \rangle^{\frac{1}{2}}$ and $\langle r^{2}_{14-23} \rangle^{\frac{1}{2}}$ describe the RMS spatial separation between the mass center of the charm quark-diquark $c[q_{1}q_{2}]$ and the mass center of the charm quark-antiquark $c\bar{q}$.

Further, since the two charm quarks $c$ are identical particles, according to symmetry, we have the following relationships: $\langle r^{2}_{13} \rangle^{\frac{1}{2}}=\langle r^{2}_{23} \rangle^{\frac{1}{2}}$, $\langle r^{2}_{14} \rangle^{\frac{1}{2}}=\langle r^{2}_{24} \rangle^{\frac{1}{2}}$, and $\langle r^{2}_{13-24} \rangle^{\frac{1}{2}}=\langle r^{2}_{14-23} \rangle^{\frac{1}{2}}$. 
Moreover, the RMS radius plays a pivotal role in differentiating between compact pentaquark states and hadronic molecular states.
More precisely, the RMS radius of a hadronic molecular state is typically around several fm, with negligible spatial overlap among its hadronic constituents;
in contrast, a compact pentaquark state displays substantial spatial overlap \cite{Luo:2022cun}.
Leveraging this distinction, we can probe the system’s spatial nature by comparing RMS spatial separation:
if the RMS spatial separation between the mass center of the charm quark-diquark $c[q_{1}q_{2}]$ and the mass center of the charm quark-antiquark $c\bar{q}$ ($\langle r^{2}_{13-24} \rangle^{\frac{1}{2}}$, $\langle r^{2}_{14-23} \rangle^{\frac{1}{2}}$)
are comparable to or even smaller than the intra-pair (anti)quark distances ($\langle r^{2}_{12} \rangle^{\frac{1}{2}}$,$\langle r^{2}_{13} \rangle^{\frac{1}{2}}$, $\langle r^{2}_{23} \rangle^{\frac{1}{2}}$, $\langle r^{2}_{14} \rangle^{\frac{1}{2}}$, $\langle r^{2}_{24} \rangle^{\frac{1}{2}}$, $\langle r^{2}_{34} \rangle^{\frac{1}{2}}$) as well as the RMS spatial separation between the mass centers of the charm quark pair ($cc$) and the $[q_{1}q_{2}]\bar{q}$ substructure ($\langle r^{2}_{12-34} \rangle^{\frac{1}{2}}$), 
which indicates strong spatial overlap between the $c[qq]$ and the $c\bar{q}$ substructures. 
Such a configuration supports the interpretation of a compact pentaquark state, as opposed to a loosely bound molecular system.

\subsection{Rearrangement decay width}\label{sec233}

\begin{figure}[t]
\includegraphics[width=0.98\linewidth]{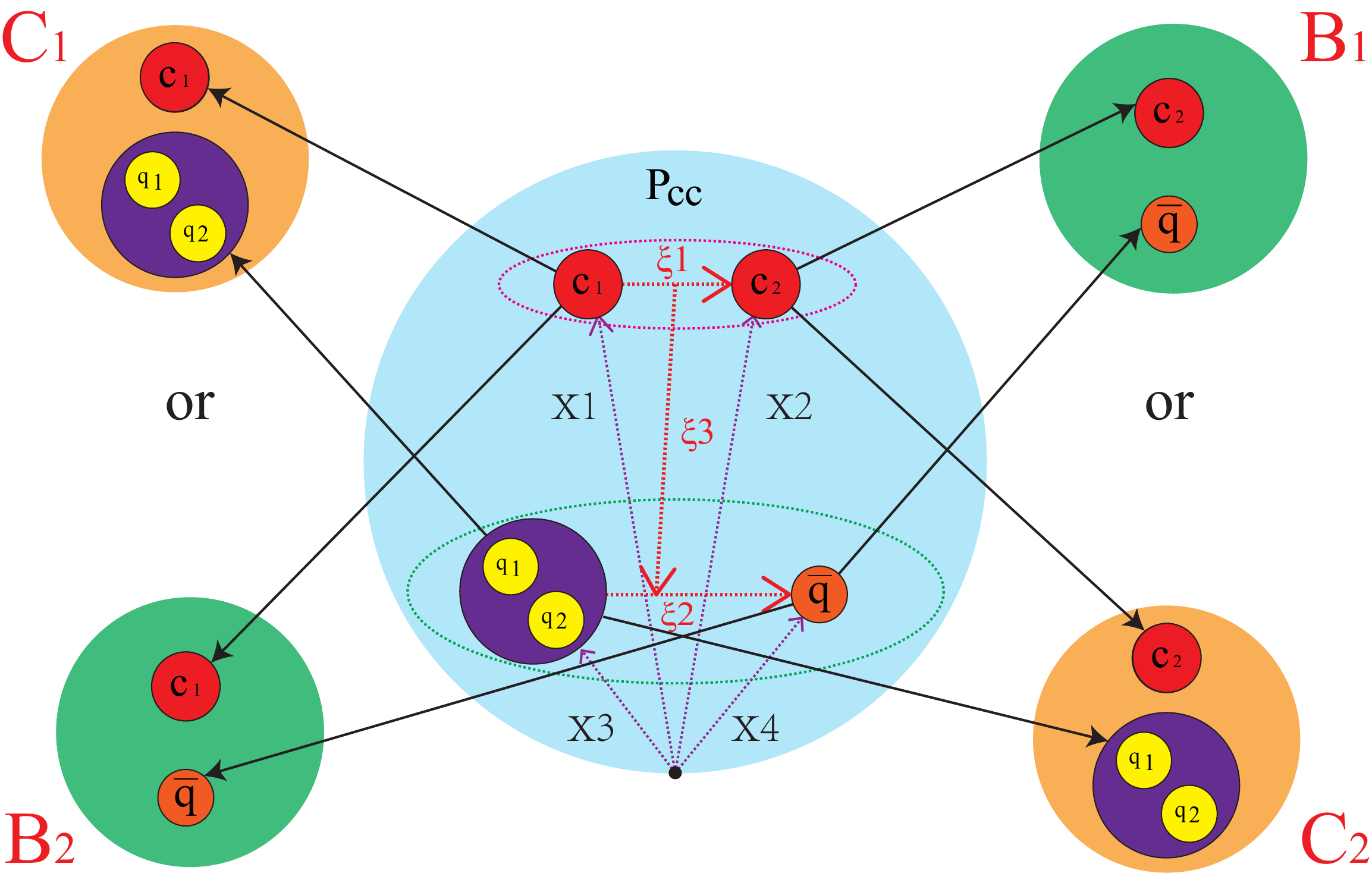}
\caption{
Spatial coordinates defined for the $P_{cc}$ doubly-charmed pentaquark system and its
two-body decays into a meson-baryon $BC$ final state via quark rearrangement. 
Here, the $BC$ final state can form via two quark rearrangement pathways: $B_{1}C_{1}$ ($[c_{2}\bar{q}][c_{1}(q_{1}q_{2})]$) and $B_{2}C_{2}$ ($[c_{1}\bar{q}][c_{2}(q_{1}q_{2})]$), as illustrated in the figure.
}\label{fig2}
\end{figure}

Besides the mass spectra and RMS radii, 
we also calculate the rearrangement decay widths of doubly-charmed pentaquarks within the framework of the quark-interchange model.
In the [$c$-$c$]-[diquark-antiquark] configuration, their dominant decay mode is $cc[qq]\bar{q}\to c[qq]+c\bar{q}$ when corresponding phase space is allowed.
Alternatively, the doubly-charmed pentaquarks can also decay to a doubly-charmed baryon ($ccq$) and a light meson ($q\bar{q}$), however this decay mode is significantly suppressed relative to the aforementioned mode, because of the diquark $[qq]$ dissociation constraints.
Meanwhile, the three-body decays are largely suppressed compared with the two-body decays due to the phase space suppression.
Thus, we do not consider them in this work.
Compared to strong decay, the contributions from radiative and weak decays are negligible.
Thus, in this paper, we focus exclusively on 
the $cc[qq]\bar{q}\to c[qq]+c\bar{q}$ decay mode, as illustrated in Fig \ref{fig2}.
Accordingly, we present all the partial widths and the total width of each state in this decay mode. 
Due to the neglect of other contributing factors, 
the actual total width of the doubly-charmed pentaquarks should be slightly larger than our theoretical results.

In the quark-interchange model \cite{Barnes:1991em,Wong:2001td},
the (di)quark-(anti)quark interactions are considered to be the sources of the fall-apart decays of pentaquarks via the quark rearrangement.
This model has been successfully applied to the rearrangement decays of many exotic states, for example, such as the 
$X(3872)$ \cite{Zhou:2019swr}, 
$X(4630)$ \cite{Yang:2021sue}, 
$X(2900)$ \cite{Wang:2020prk,Liu:2022hbk}, 
$Z_{c}$ and $Z_{b}$ states \cite{Wang:2018pwi,Xiao:2019spy},
hidden and double charm-strange tetraquark \cite{Liu:2024fnh}, 
fully-charm tetraquark \cite{liu:2020eha},
hidden-charm pentaquark $P_{c}$ states \cite{Wang:2019spc},
hidden-charm pentaquarks with triple strangeness \cite{Wang:2021hql}, fully-charm pentaquarks \cite{Liang:2024met},
and doubly-charmed hexaquarks \cite{An:2025rjv}.
Moreover, this model has achieved a reliable description of the low-energy $S$-wave phase shift for the $I=2$ $\pi\pi$ scattering at the quark level \cite{Barnes:2000hu}.

Here, the partial decay width $\Gamma$ for the decay process 
$A\to BC$ is given by
\begin{eqnarray}\label{width}
\Gamma=\frac{1}{(2J_{A}+1)}\frac{|\vec{P}_{B}|}{32\pi^{2}M^{2}_{A}}\int d\Omega|\mathcal{M}(A \to BC)|^{2},
\end{eqnarray}
where $A$ stands for the initial doubly-charmed pentaquarks, 
$B$ and $C$ stand for the final single-charmed meson and single-charmed baryon, respectively (see Fig \ref{fig2}).
Meanwhile, $\vec{P}_{B}$ is the three-vector momentum of the final state $B$ or $C$ in the initial-hadron-rest frame, and $M_{A}$ is the mass of the initial doubly-charmed pentaquarks.
The $\mathcal{M}(A \to BC)$ stands for decay amplitude, which is described by
\begin{eqnarray}\label{amp}
\mathcal{M}(A \to BC)=
-(2\pi)^{3/2}\sqrt{2M_{A}}\sqrt{2E_{B}}\sqrt{2E_{C}}\times T,
\end{eqnarray}
where $E_{B}$ and $E_{C}$ are the energies of the final states $B$ and $C$, respectively.
The $T$-matrix is expressed as:
\begin{eqnarray}\label{Eq:T1}
T&=&\langle\Psi^{B}\Psi^{C}|\sum_{i<j}V_{ij}|\Psi^{A}\rangle\nonumber\\
&=&\langle\Psi^{B}\Psi^{C}|\sum_{i<j}V_{ij}|\Psi^{A}_{(cc)}\Psi^{A}_{([qq]\bar{q})}\Psi^{A}_{(cc)-([qq]\bar{q})}\rangle.\quad
\end{eqnarray}
where $\Psi^{A}$, $\Psi^{B}$, and $\Psi^{C}$ represent the total wave functions of the initial doubly-charmed pentaquarks, final meson, and final baryon, respectively.
$V_{ij}$ is the potential as listed in Eq.(\ref{Eq2}).
The $T$-matrix in the momentum space can be written as:
\begin{eqnarray}\label{Eq:T2}
T=\frac{1}{(2\pi)^{3}}\int d\vec{P}_{\alpha}V_{{\rm eff}}(\vec{P}_{\alpha},\vec{P}_{B})\Psi^{A}_{(cc)-([qq]\bar{q})}(\vec{P}_{\alpha}).
\end{eqnarray}
Here, the effective potential $V_{{\rm eff}}(\vec{P}_{\alpha},\vec{P}_{B})$, which is a function of the initial and final relative momenta $\vec{P}_{\alpha}$ and $\vec{P}_{B}$, is calculated as the overlaps of the wave functions with the potential between the initial and final states.

\begin{figure*}[htbp]
\includegraphics[width=0.98\linewidth]{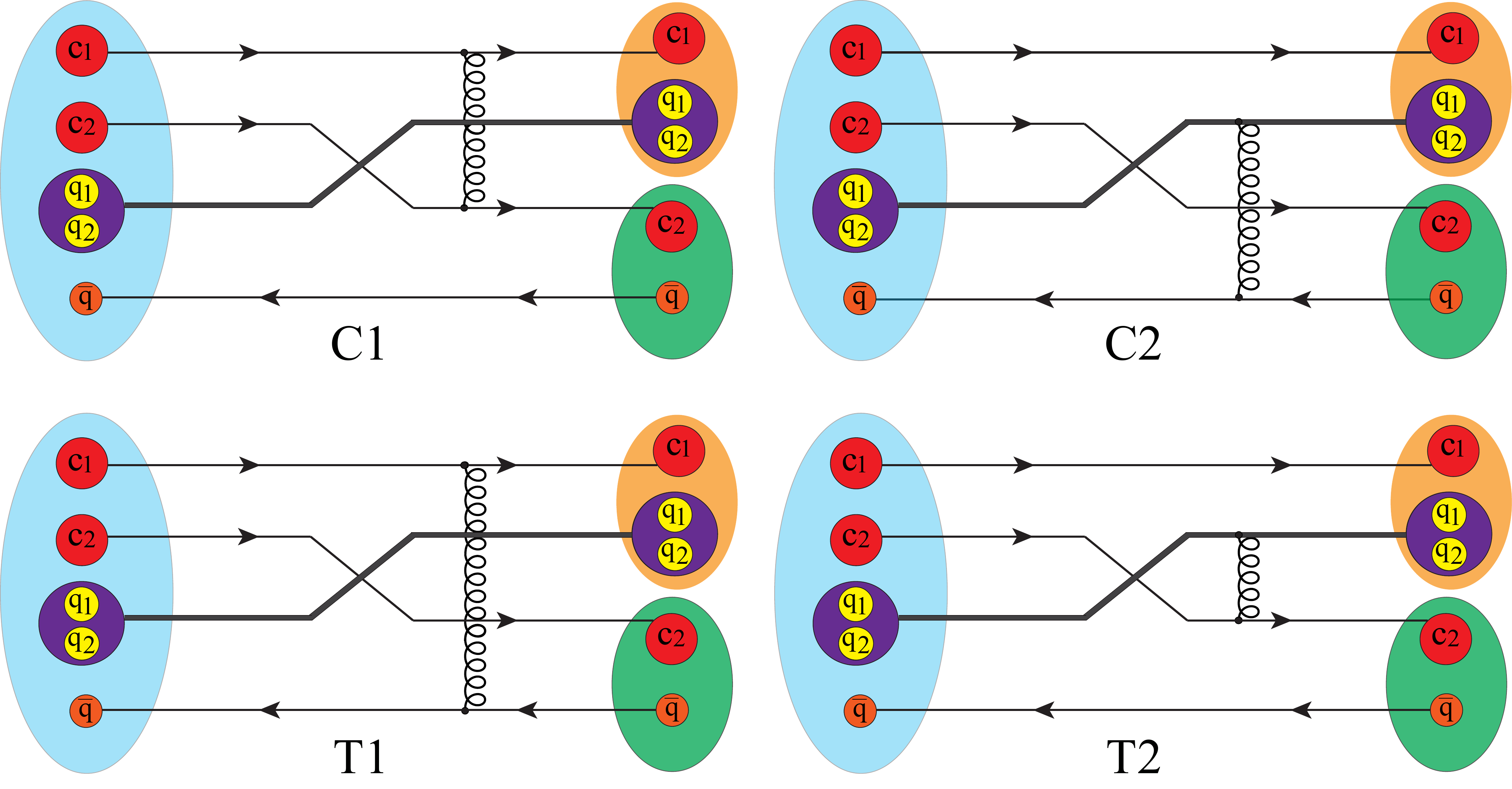}
\caption{
The quark-interchange diagrams for $P_{cc}$ decaying into meson-baryon final state at the quark level.
The curve line denotes the (di)quark–(anti)quark interactions.
}\label{fig3}
\end{figure*}

The $V_{{\rm eff}}(\vec{P}_{\alpha},\vec{P}_{B})$ 
combines interaction contributions from four diagrams $C_{1}$, $C_{2}$, $T_{1}$, and $T_{2}$ in Fig.~\ref{fig3}.
For each diagram, the $V_{eff}(\vec{P}_{\alpha},\vec{P}_{B})$ is factorized into the product of factors
\begin{eqnarray}
V_{{\rm eff}}(\vec{P}_{\alpha},\vec{P}_{B})
=I_{{\rm flavor}}I_{{\rm color}}I_{{\rm spin-space}}.
\end{eqnarray}
Here, $I$ with the subscripts flavor, color, and spin-space represent the overlaps of the initial and final wave functions in the corresponding space.

Firstly, the flavor factor $I_{{\rm flavor}}$ is simply unity for all diagrams considered in our paper.
Then, the color factor $I_{{\rm color}}$ is:
\begin{eqnarray}\label{eq:icolor}
I_{{\rm color}}=\langle\psi_{B}\psi_{C}|\frac{\lambda^{c}_{i}}{2}.\frac{\lambda^{c}_{j}}{2}|\psi^{\bar{3}_{c}(6_{c})}_{A}(cc)\psi^{3_{c}(\bar{6}_{c})}_{A}([qq]\bar{q})\rangle.\quad
\end{eqnarray}
The numerical results in different diagrams in Fig.\ref{fig3} are listed in Table \ref{icolor}.
As for the spin-space factor $I_{{\rm spin-space}}$, the spin and space factors can be decoupled in the $S$-wave decay process.
The spin factor $I_{{\rm spin}}$ is:
\begin{eqnarray}\label{ispin}
I_{{\rm spin}}=\langle[\chi^{B}_{s_{3}}\chi^{C}_{s_{4}}]_{S'} |\hat{\mathcal{O}}_{s}|[\chi(cc)^{A}_{s_{1}}\chi([qq]\bar{q})^{A}_{s_{2}}]_{S}\rangle,
\end{eqnarray}
where $s_{1}$ and $s_{2}$ represent the spins of initial
$cc$ pair and $[qq]\bar{q}$ substructure,
$s_{3}$ and $s_{4}$ represent the spins of the final meson and baryon, 
$S$ and $S'$ represent the total spin of initial and final state.
The $\hat{\mathcal{O}}_{s}$ represents the spin operator, which takes $\textbf{1}$ for the Coulomb and linear confinement potentials, and $\frac{\sigma_{i}}{2}\cdot\frac{\sigma_{j}}{2}$ for the hyperfine potential.
Finally, the space factor $I_{{\rm space}}$ is:
\begin{eqnarray}\label{ispace}
&&I_{{\rm space}}=\langle\Psi^{B}\Psi^{C}|\hat{\mathcal{O}}_{q}|\Psi^{A}_{(cc)}\Psi^{A}_{([qq]\bar{q})}\rangle \nonumber\\
&&=\int\int d\textbf{k}_{1}d\textbf{k}_{2}\Psi^{B}(k_{B}+K_{B})\Psi^{C}(k_{C}+K_{C})\hat{\mathcal{O}}_{q}(k_{1}-k_{2})\nonumber\\
&&\quad \Psi^{*A}_{(cc)}(k_{\alpha}+K_{\alpha})\Psi^{*A}_{([qq]\bar{q})}(k_{\beta}+K_{\beta}),
\end{eqnarray}
where $\hat{\mathcal{O}}_{q}$ represents the spatial operator, which takes $1/q^{2}$, $1/q^{4}$, and $\exp[-q^{2}]$ for the Coulomb, linear confinement, and hyperfine potentials, respectively.
The $\vec{k}_{1}$ ($\vec{k}_{2}$) is the initial (final) three-momenta of the scattered constituent.
We denote the three-momenta of the final-state meson $B$, the final-state baryon $C$, the ($cc$) component, and the ($[qq]\bar{q}$) component of the initial doubly-charmed pentaquark state as $\vec{P}_{B}$,  $\vec{P}_{C}$, $\vec{P}_{\alpha}$, and $\vec{P}_{\beta}$, respectively.
In the center-of-mass frame, the three-momenta satisfy $\vec{P}_{B}=-\vec{P}_{C}$ and $\vec{P}_{\alpha}=-\vec{P}_{\beta}$.

Based on the above relations, the corresponding momenta $\vec{k}_{i}$ and $\vec{K}_{i}$ ($i=B, C, \alpha, \beta$) for the four quark exchange diagrams in Fig.~\ref{fig3} are derived and presented in Table \ref{icolor} in terms of $\vec{P}_{B}$, $\vec{P}_{\alpha}$, $\vec{k}_{1}$, and $\vec{k}_{2}$.
In Table \ref{icolor}, the constituent quark mass-dependent functions $f_{i}$ ($i= \alpha, \beta, B, C$) are expressed as:
\begin{eqnarray}
f_{\alpha}&=&\frac{m_{c}}{m_{c}+m_{c}}=\frac{1}{2}, \quad f_{\beta}=\frac{m_{[q_{1}q_{2}]}}{m_{[q_{1}q_{2}]}+m_{\bar{q}}}, \nonumber \\
f_{B}&=&\frac{m_{[q_{1}q_{2}]}}{m_{c}+m_{[q_{1}q_{2}]}}, \quad \quad
f_{C}=\frac{m_{\bar{q}}}{m_{c}+m_{\bar{q}}}.
\end{eqnarray}
Here, $m_{c}$, $m_{[q_{1}q_{2}]}$, and $m_{\bar{q}}$ represent the masses of charm quark, light diquark, and antiquark, respectively, with their values presented in Table \ref{para}.
The detailed derivation of the integral simplification for Eq. (\ref{ispace}) can be found in Refs.~\cite{Barnes:1991em,Wong:2001td}.

Finally, utilizing the calculated decay amplitude $\mathcal{M}(A \to BC)$, we calculate the rearrangement decay widths via Eq.~(\ref{width}) and tabulate them in Tables \ref{ccnnn}-\ref{ccnsn}.

\begin{table*}[t]
\centering \caption{ The color matrix element $I_{\rm color}=\langle\frac{\lambda^{c}_{i}}{2}.\frac{\lambda^{c}_{j}}{2}\rangle$ and the momentum substitutions in $I_{\rm space}$ for different scattering diagrams.
}\label{icolor}
\begin{lrbox}{\tablebox}
\renewcommand\arraystretch{2}
\renewcommand\tabcolsep{3.5 pt}
\begin{tabular}{c!{\vrule width 0.75pt}cc!{\vrule width 0.75pt}c|c!{\vrule width 0.75pt}c|c!{\vrule width 0.75pt}c|c!{\vrule width 0.75pt}c|c}
\toprule[1.50pt]
\toprule[0.50pt]
\multirow{2}*{Diagram}&\multicolumn{2}{c!{\vrule width 0.75pt}}{$\langle\frac{\lambda^{c}_{i}}{2}.\frac{\lambda^{c}_{j}}{2}\rangle$}&\multicolumn{8}{c}{momentum substitutions}\\
\Xcline{2-11}{0.75pt}
&$(cc)^{\bar{3}_{c}}([qq]\bar{q})^{3_{c}}$&$(cc)^{6_{c}}([qq]\bar{q})^{\bar{6}_{c}}$&$\vec{k}_{\alpha}$&$\vec{K}_{\alpha}$&$\vec{k}_{\beta}$&$\vec{K}_{\beta}$&$\vec{k}_{B}$&$\vec{K}_{B}$&$\vec{k}_{C}$&$\vec{K}_{C}$\\
\Xcline{1-11}{0.75pt}
$C_{1}$&$\frac{2\sqrt{3}}{9}$&$\frac{\sqrt{6}}{9}$&$\vec{k}_{1}$&$-f_{\alpha}\vec{P}_{\alpha}$&$\vec{k}_{2}$&$(1-f_{\beta})\vec{P}_{\alpha}-\vec{P}_{B}$
&$\vec{k}_{2}$&$-f_{B}\vec{P}_{B}$&$\vec{k}_{2}$&$-f_{C}\vec{P}_{B}-\vec{P}_{\alpha}$\\
$C_{2}$&$\frac{2\sqrt{3}}{9}$&$\frac{\sqrt{6}}{9}$&$\vec{k}_{1}$&$-f_{\alpha}\vec{P}_{\alpha}$&$\vec{k}_{2}$&$-f_{\beta}\vec{P}_{\alpha}+\vec{P}_{B}$
&$\vec{k}_{1}$&$f_{B}\vec{P}_{B}-\vec{P}_{\alpha}$&$\vec{k}_{1}$&$f_{\beta}\vec{P}_{\alpha}$\\
$T_{1}$&$-\frac{2\sqrt{3}}{9}$&$-\frac{\sqrt{6}}{9}$&$\vec{k}_{1}$&$-f_{\alpha}\vec{P}_{\alpha}$&$\vec{k}_{2}$&$(1-f_{\beta})\vec{P}_{\alpha}-\vec{P}_{B}$&$\vec{k}_{2}$&$-f_{\alpha}\vec{P}_{\alpha}$&$\vec{k}_{1}$&$-f_{C}\vec{P}_{B}-\vec{P}_{\alpha}$\\
$T_{2}$&$-\frac{2\sqrt{3}}{9}$&$-\frac{\sqrt{6}}{9}$&$\vec{k}_{1}$&$-f_{\alpha}\vec{P}_{\alpha}$&$\vec{k}_{2}$&$-f_{\beta}\vec{P}_{\alpha}+\vec{P}_{B}$&
$\vec{k}_{1}$&$f_{B}\vec{P}_{B}-\vec{P}_{\alpha}$&$\vec{k}_{2}$&$f_{\beta}\vec{P}_{\alpha}$\\
\toprule[0.50pt]
\toprule[1.50pt]
\end{tabular}
\end{lrbox}\scalebox{1.05}{\usebox{\tablebox}}
\end{table*}

\section{Results and discussions}\label{sec3}

In this section, we present our calculation results of the mass spectra, internal mass contributions, RMS radii, and rearrangement decay properties in Tables \ref{ccnnn}-\ref{ccnsn}.

Firstly, the masses of the doubly-charmed pentaquarks are theoretically predicted to lie in the range of 4.7-5.4 GeV.
Subsequently, the internal mass contributions indicate that the kinetic energy $\langle T \rangle$ and confinement potential $\langle V^{\rm Con} \rangle$ have the same order of magnitude. 
Furthermore, the internal mass contributions indicate that since the off-diagonal elements of the hyperfine interaction potential $\langle V^{\rm SS} \rangle$ matrix are non-zero, it plays an important role in the mixing of different $|(cc)^{\rm color_{1}}_{\rm spin_{1}}[(q_{1}q_{2})^{\bar{3}_{c}}_{\rm spin_{2}}\bar{q}_{\frac{1}{2}}]^{\rm color_{2}}_{\rm spin_{3}}\rangle_{\rm spin_{4}}$ configurations.
This color-spin configuration mixing 
results in notable mass shifts and larger mass gaps of the physical states compared with their pre-mixing counterparts.
After considering the configuration mixing, we obtain the masses of corresponding physical states and summarize them in Tables \ref{ccnnn}-\ref{ccnsn}.

Since all predicted physical states lie above their corresponding baryon-meson thresholds, 
we further evaluate their rearrangement decay properties and present the partial widths and total widths of each state calculated via Eq.(\ref{width}).
According to Tables \ref{ccnnn}-\ref{ccnsn}, the total widths of the doubly-charmed pentaquarks are predicted to be in the range of 15-70 MeV.
Notably, their rearrangement decay properties can also play a crucial role in distinguishing partner pentaquarks with similar masses.
Meanwhile, we have also extracted the partial width ratios, thereby allowing us to propose that relevant experimental collaborations search for the pentaquarks in the corresponding meson-baryon decay final states.

To enhance clarity, 
the relative mass positions of each state, total decay widths, partial width ratios, and corresponding rearrangement decay channels—all derived from the numerical results presented in Tables \ref{ccnnn}–\ref{ccnsn}—are illustrated in Figs. \ref{fig-ccnnN}–\ref{fig-ccnsN}.
For convenience, we also label all possible spin (isospin) quantum numbers of the rearrangement decay channels with subscripts (superscripts).
When the isospin(spin) of an initial pentaquark is equal to a number in the subscript(superscript) of a baryon-meson final state, 
the pentaquark can decay into that baryon-meson channel, which is allowed by the isospin(angular momentum) conservation.
As shown in the preceding figures or the decay properties summarized in corresponding Tables, 
there is no stable state in the doubly-charmed pentaquark system.
Conversely, all such states are unstable, 
undergoing rearrangement decay to a singly-charmed meson and a singly-charmed baryon as final states.
The primary reason is that the pairwise attractive interactions provided by $\langle V^{\rm Con} \rangle$ are significantly weaker than corresponding singly-charmed baryon and meson. 
Consequently, their masses lie above the thresholds of the corresponding decay final states.

In addition to the mass spectra and rearrangement decay properties, 
we also present the corresponding RMS radii calculated via Eq. (\ref{rms}). 
As illustrated in Tables \ref{ccnnn}-\ref{ccnsn}, 
most RMS radii fall within the range of 1.1–1.6 fm, 
exhibiting roughly the same order of magnitude.
Meanwhile, our calculations yield the following relations: $\langle r^{2}_{13} \rangle^{\frac{1}{2}}=\langle r^{2}_{23} \rangle^{\frac{1}{2}}$, $\langle r^{2}_{14} \rangle^{\frac{1}{2}}=\langle r^{2}_{24} \rangle^{\frac{1}{2}}$, and $\langle r^{2}_{13-24} \rangle^{\frac{1}{2}}=\langle r^{2}_{14-23} \rangle^{\frac{1}{2}}$, which are in full agreement with our symmetry analysis presented in Subsec.\ref{sec232}.
If the configuration is molecular, 
$\langle r^{2}_{13-24} \rangle^{\frac{1}{2}}$ and 
$\langle r^{2}_{14-23} \rangle^{\frac{1}{2}}$ should be significantly larger than the other RMS radii—especially 
$\langle r^{2}_{12-34} \rangle^{\frac{1}{2}}$
and on the order of several femtometers, 
with the spatial overlap components ($\langle r^{2}_{13} \rangle^{\frac{1}{2}}$, $\langle r^{2}_{14} \rangle^{\frac{1}{2}}$, $\langle r^{2}_{23} \rangle^{\frac{1}{2}}$, and $\langle r^{2}_{24} \rangle^{\frac{1}{2}}$) being negligible in comparison.
Thus, the results of our calculations align with the expectations for the compact pentaquark configuration.

For clarity in subsequent discussion, we adopt the notation $\rm P_{content}(I,J^{P},Mass)$ to denote a specific pentaquark state.

\subsection{The $cc[nn]\bar{n}$ and $cc[nn]\bar{s}$ subsystems}

\begin{table*}[t]
\centering
\caption{
The numerical results of the mass spectrum, the mass contributions of each Hamiltonian part (in MeV), the root-mean-square radii (in fm), and the partial and total decay widths of the rearrangement decay (in MeV) 
for the $cc[nn]\bar{n}$ and $cc[nn]\bar{s}$ pentaquarks. 
}\label{ccnnn}
\begin{lrbox}{\tablebox}
\renewcommand\arraystretch{2.1}
\renewcommand\tabcolsep{0.75 pt}
\begin{tabular}{ccc|ccc|cccccc|rrrrrr|r}
\toprule[1.50pt]
\toprule[0.50pt]
\multicolumn{3}{l|}{$cc[nn]\bar{n}$}&
\multicolumn{3}{c|}{Internal contribution}& 
\multicolumn{6}{c|}{RMS Radius}&\multicolumn{7}{c}{Rearrangement decay properties}\\
\Xcline{4-19}{0.3pt}
\multirow{2}*{$I[J^{P}]$}&\multirow{2}*{Configuration}&\multirow{2}*{Mass}&\multirow{2}*{$\langle T \rangle$}
&\multirow{2}*{$\langle V^{\rm Con} \rangle$}
&\multirow{2}*{$\langle V^{\rm SS} \rangle$}
&\multirow{2}*{$R_{12}$}&\multirow{2}*{$R_{34}$}
&$R_{13}$&$R_{14}$
&\multirow{2}*{$R_{12-34}$}
&$R_{13-24}$
&\multirow{2}*{$\Sigma^{*}_{c}D^{*}$}
&\multirow{2}*{$\Sigma^{*}_{c}D$}
&\multirow{2}*{$\Sigma_{c}D^{*}$}
&\multirow{2}*{$\Sigma_{c}D$}
&\multirow{2}*{$\Lambda_{c}D^{*}$}
&\multirow{2}*{$\Lambda_{c}D$}
&\multicolumn{1}{c}{\multirow{2}*{$\Gamma_{sum}$}}\\
&&&&&&&&$R_{23}$&$R_{24}$&&$R_{14-23}$&&&&&&&\\
\bottomrule[1.00pt]
\multirow{1}*{$\frac{3}{2}(\frac{1}{2})[5/2^{-}]$}&\multirow{1}*{$
|(cc)^{I=0,\bar{3}_{c}}_{s=1}[(nn)^{I=1,\bar{3}_{c}}_{s=1}\bar{n}]^{I=3/2,3_{c}}_{s=3/2}\rangle^{I=3/2}_{s=5/2}$}&5100.6&1158.6&-829.1&47.1&1.16&1.50&1.24&1.53&1.37&1.26&12.6&&&&&&12.6\\
\multirow{3}*{$\frac{3}{2}(\frac{1}{2})[3/2^{-}]$}&
\multirow{3}*{$\begin{pmatrix}
|(cc)^{I=0,\bar{3}_{c}}_{s=1}[(nn)^{I=1,\bar{3}_{c}}_{s=1}\bar{n}]^{I=3/2,3_{c}}_{s=3/2}\rangle^{I=3/2}_{s=3/2}
\\
|(cc)^{I=0,\bar{3}_{c}}_{s=1}[(nn)^{I=1,\bar{3}_{c}}_{s=1}\bar{n}]^{I=3/2,3_{c}}_{s=1/2}\rangle^{I=3/2}_{s=3/2}
\\
|(cc)^{I=0,6_{c}}_{s=0}[(nn)^{I=1,\bar{3}_{c}}_{s=1}\bar{n}]^{I=3/2,\bar{6}_{c}}_{s=3/2}\rangle^{I=3/2}_{s=3/2}
\\
\end{pmatrix}$}
&
\multirow{3}*{$\begin{pmatrix}
5140.8\\5070.7\\4962.0
\end{pmatrix}$}
&1186.9&-806.3&-5.0&1.36&1.48&1.20&1.50&1.20&1.40&21.5&9.1&13.3&&&&43.9
\\
&&&1181.2&-852.0&-5.8&1.16&1.49&1.23&1.53&1.36&1.26&2.2&3.0&5.6&&&&10.8\\
&&&1207.6&-877.8&-39.0&1.16&1.47&1.23&1.51&1.36&1.25&11.9&4.8&5.7&&&&22.4\\
\multirow{3}*{$\frac{3}{2}(\frac{1}{2})[1/2^{-}]$}&
\multirow{3}*{$\begin{pmatrix}
|(cc)^{I=0,\bar{3}_{c}}_{s=1}[(nn)^{I=1,\bar{3}_{c}}_{s=1}\bar{n}]^{I=3/2,3_{c}}_{s=3/2}\rangle^{I=3/2}_{s=1/2}
\\
|(cc)^{I=0,\bar{3}_{c}}_{s=1}[(nn)^{I=1,\bar{3}_{c}}_{s=1}\bar{n}]^{I=3/2,3_{c}}_{s=1/2}\rangle^{I=3/2}_{s=1/2}
\\
|(cc)^{I=0,6_{c}}_{s=0}[(nn)^{I=1,\bar{3}_{c}}_{s=1}\bar{n}]^{I=3/2,\bar{6}_{c}}_{s=1/2}\rangle^{I=3/2}_{s=1/2}
\\
\end{pmatrix}$}
&
\multirow{3}*{$\begin{pmatrix}
5167.5\\5038.7\\4947.2
\end{pmatrix}$}
&1170.5&-789.8&22.2&1.36&1.49&1.20&1.51&1.20&1.40&31.9&&3.9&7.2&&&42.9\\
&&&1198.6&-868.9&-23.7&1.16&1.48&1.24&1.52&1.36&1.25&3.3&&8.3&1.4&&&13.0\\
&&&1217.0&-887.2&-57.4&1.16&1.48&1.22&1.51&1.35&1.25&9.0&&4.3&5.3&&&18.6\\
\toprule[0.05pt]
\multirow{1}*{$1/2[3/2^{-}]$}&
\multirow{1}*{$
|(cc)^{I=0,\bar{3}_{c}}_{s=1}[(nn)^{I=0,\bar{3}_{c}}_{s=0}\bar{n}]^{I=1/2,3_{c}}_{s=1/2}\rangle^{I=1/2}_{s=3/2}$}
&4799.2&1181.5&-791.6&24.9&1.17&1.51&1.28&1.53&1.40&1.27&&&&&7.1&&7.1\\
\multirow{2}*{$1/2[1/2^{-}]$}&
\multirow{2}*{$\begin{pmatrix}
|(cc)^{I=0,\bar{3}_{c}}_{s=1}[(nn)^{I=0,\bar{3}_{c}}_{s=0}\bar{n}]^{I=1/2,3_{c}}_{s=1/2}\rangle^{I=1/2}_{s=1/2}
\\
|(cc)^{I=0,6_{c}}_{s=0}[(nn)^{I=0,\bar{3}_{c}}_{s=0}\bar{n}]^{I=3/2,\bar{6}_{c}}_{s=1/2}\rangle^{I=1/2}_{s=1/2}
\\
\end{pmatrix}$}
&
\multirow{2}*{$\begin{pmatrix}
4883.7\\4694.7
\end{pmatrix}$}&1190.6&-737.8&4.1&1.37&1.51&1.24&1.50&1.24&1.44&&&&&35.0&8.4&43.4\\
&&&1216.8&-826.7&-37.7&1.16&1.50&1.26&1.52&1.39&1.26&&&&&12.1&4.9&17.0\\
\Xcline{1-19}{1pt}
\multicolumn{3}{l|}{$cc[nn]\bar{s}$}&\multirow{2}*{$\langle T \rangle$}
&\multirow{2}*{$\langle V^{\rm Con} \rangle$}
&\multirow{2}*{$\langle V^{\rm SS} \rangle$}
&\multirow{2}*{$R_{12}$}&\multirow{2}*{$R_{34}$}
&$R_{13}$&$R_{14}$
&\multirow{2}*{$R_{12-34}$}
&$R_{13-24}$
&\multirow{2}*{$\Sigma^{*}_{c}D_{s}^{*}$}
&\multirow{2}*{$\Sigma^{*}_{c}D_{s}$}
&\multirow{2}*{$\Sigma_{c}D_{s}^{*}$}
&\multirow{2}*{$\Sigma_{c}D_{s}$}
&\multirow{2}*{$\Lambda_{c}D_{s}^{*}$}
&\multirow{2}*{$\Lambda_{c}D_{s}$}
&\multicolumn{1}{c}{\multirow{2}*{$\Gamma_{sum}$}}\\
\multirow{1}*{$I[J^{P}]$}&\multirow{1}*{Configuration}&\multirow{1}*{Mass}&&&&&&$R_{23}$&$R_{24}$&&$R_{14-23}$&&&&&&&\\
\Xcline{1-19}{0.3pt}
\multirow{1}*{$1[5/2^{-}]$}&
\multirow{1}*{$
|(cc)^{I=0,\bar{3}_{c}}_{s=1}[(nn)^{I=1,\bar{3}_{c}}_{s=1}\bar{s}]^{I=1,3_{c}}_{s=3/2}\rangle^{I=1}_{s=5/2}
$}
&5143.7&1097.6&-1046.2&41.2&1.16&1.37&1.24&1.38&1.33&1.24&10.1&&&&&&10.1\\
\multirow{3}*{$1[3/2^{-}]$}&
\multirow{3}*{$\begin{pmatrix}
|(cc)^{I=0,\bar{3}_{c}}_{s=1}[(nn)^{I=1,\bar{3}_{c}}_{s=1}\bar{s}]^{I=1,3_{c}}_{s=3/2}\rangle^{I=1}_{s=3/2}
\\
|(cc)^{I=0,\bar{3}_{c}}_{s=1}[(nn)^{I=1,\bar{3}_{c}}_{s=1}\bar{s}]^{I=1,3_{c}}_{s=1/2}\rangle^{I=1}_{s=3/2}
\\
|(cc)^{I=0,6_{c}}_{s=0}[(nn)^{I=1,\bar{3}_{c}}_{s=1}\bar{s}]^{I=1,\bar{6}_{c}}_{s=3/2}\rangle^{I=1}_{s=3/2}
\\
\end{pmatrix}$}
&\multirow{3}*{$\begin{pmatrix}
5139.0\\5115.1\\5013.6
\end{pmatrix}$}
&1114.2&-1072.3&-3.4&1.33&1.36&1.20&1.35&1.17&1.35&31.3&12.6&18.9&&&&62.7\\
&&&1118.8&-1067.7&6.5&1.15&1.37&1.23&1.37&1.32&1.23&1.6&5.1&11.1&&&&17.8\\
&&&1144.7&-1082.9&-3.3&1.15&1.35&1.23&1.36&1.32&1.22&23.2&9.4&13.6&&&&46.2\\
\multirow{3}*{$1[1/2^{-}]$}&
\multirow{3}*{$\begin{pmatrix}
|(cc)^{I=0,\bar{3}_{c}}_{s=1}[(nn)^{I=1,\bar{3}_{c}}_{s=1}\bar{s}]^{I=1,3_{c}}_{s=3/2}\rangle^{I=1}_{s=1/2}
\\
|(cc)^{I=0,\bar{3}_{c}}_{s=1}[(nn)^{I=1,\bar{3}_{c}}_{s=1}\bar{s}]^{I=1,3_{c}}_{s=1/2}\rangle^{I=1}_{s=1/2}
\\
|(cc)^{I=0,6_{c}}_{s=0}[(nn)^{I=1,\bar{3}_{c}}_{s=1}\bar{s}]^{I=0,\bar{6}_{c}}_{s=1/2}\rangle^{I=1}_{s=1/2}
\\
\end{pmatrix}$}
&\multirow{3}*{$\begin{pmatrix}
5158.2\\5076.3\\5001.7
\end{pmatrix}$}
&1098.7&-1056.6&19.6&1.34&1.37&1.20&1.36&1.17&1.35&46.9&&6.4&11.6&&&64.9\\
&&&1137.1&-1085.5&-23.2&1.16&1.23&1.35&1.37&1.32&1.23&3.4&&9.9&1.8&&&15.1\\
&&&1153.3&-1101.6&-48.8&1.15&1.35&1.22&1.36&1.31&1.22&20.6&&5.8&8.2&&&34.6\\
\toprule[0.05pt]
\multirow{1}*{$0[3/2^{-}]$}&
\multirow{1}*{$
|(cc)^{I=0,\bar{3}_{c}}_{s=1}[(nn)^{I=0,\bar{3}_{c}}_{s=0}\bar{s}]^{I=0,3_{c}}_{s=1/2}\rangle^{I=0}_{s=3/2}$}
&4849.1&1121.8&-1005.0&20.8&1.16&1.39&1.28&1.37&1.36&1.25&&&&&5.9&&5.9\\
\multirow{2}*{$0[1/2^{-}]$}&
\multirow{2}*{$\begin{pmatrix}
|(cc)^{I=0,\bar{3}_{c}}_{s=1}[(nn)^{I=0,\bar{3}_{c}}_{s=0}\bar{s}]^{I=0,3_{c}}_{s=1/2}\rangle^{I=0}_{s=1/2}
\\
|(cc)^{I=0,6_{c}}_{s=0}[(nn)^{I=0,\bar{3}_{c}}_{s=0}\bar{s}]^{I=0,\bar{6}_{c}}_{s=1/2}\rangle^{I=0}_{s=1/2}
\\
\end{pmatrix}$}
&\multirow{2}*{$\begin{pmatrix}
4881.8\\4751.1
\end{pmatrix}$}
&1118.0&-1000.2&-4.3&1.35&1.40&1.25&1.35&1.20&1.39&&&&&25.8&10.3&36.1\\
&&&1154.2&-1037.2&-29.2&1.15&1.38&1.27&1.36&1.34&1.24&&&&&39.6&11.0&50.6\\
\bottomrule[0.50pt]
\bottomrule[1.50pt]
\end{tabular}
\end{lrbox}\scalebox{0.8}{\usebox{\tablebox}}
\end{table*}

\begin{figure*}[htbp]
\begin{tabular}{c}
\includegraphics[width=440pt]{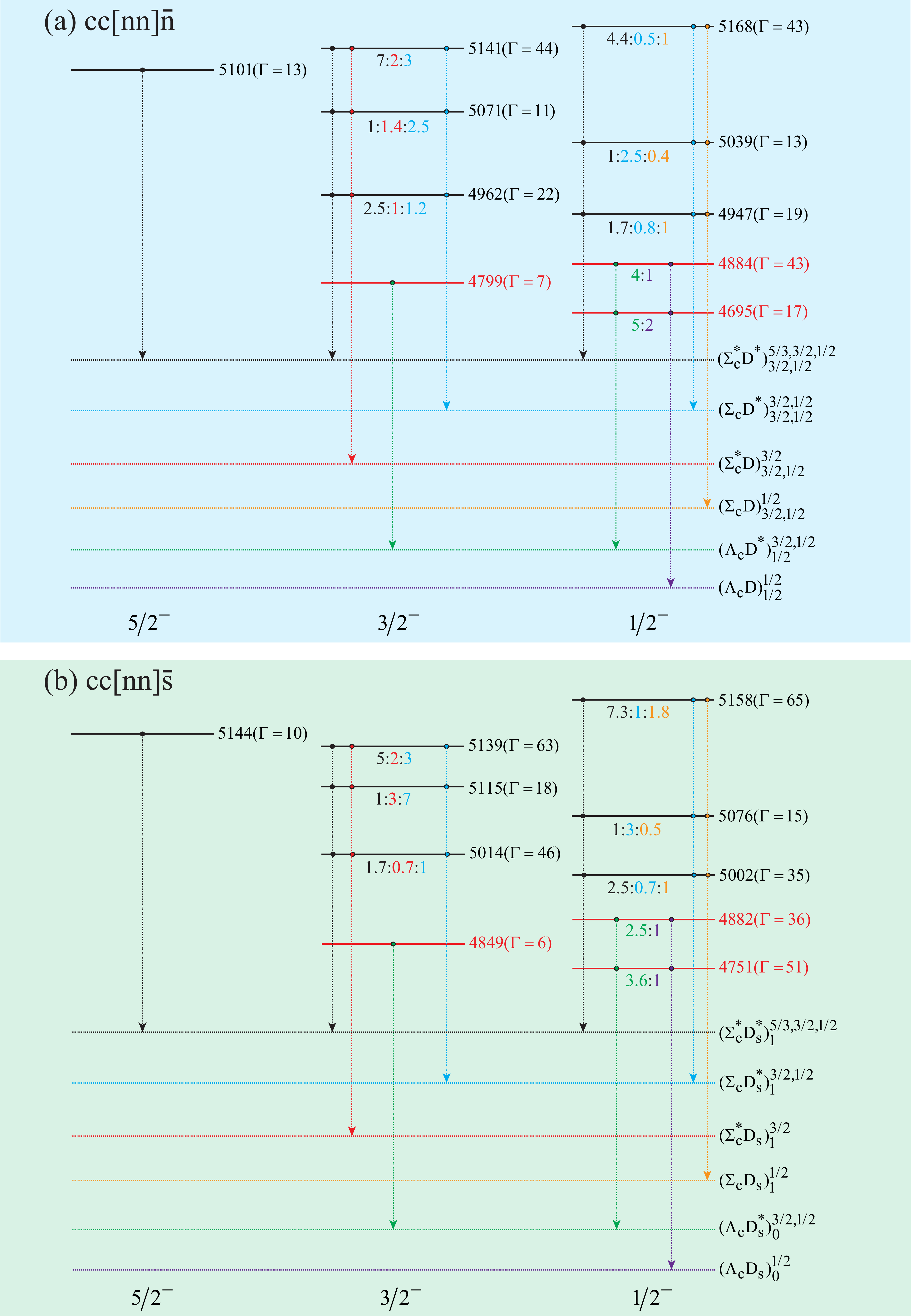}\\
\end{tabular}
\caption{
Relative positions for the $cc[nn]\bar{n}$ (a) and $cc[nn]\bar{s}$ (b) pentaquarks labeled with horizontal solid lines, e.g. $5101(\Gamma=13)$ represents the mass and total decay width of the corresponding state (units: MeV).
The numbers below the horizontal lines, e.g. $7:2:3$, represent the relative branching ratios of the corresponding state.
The black and red horizontal lines represent the pentaquarks with $I_{nn}=1$ and $0$, respectively.
The dotted lines denote various $S$-wave baryon-meson thresholds, and the superscripts (subscript) of the labels, e.g. $(\Sigma^{*}_{c}D^{*})^{5/2,3/2,1/2}_{3/2,1/2}$, represent the possible total angular momenta (isospin) of the channels.
The solid dots of different colors where the vertical dashed lines with arrows intersect the horizontal solid lines represent the allowed rearranged S-wave decay processes. 
If a vertical dashed line with an arrow intersects a horizontal solid line without a solid dot, it means that corresponding decay process is forbidden for relevant state.
}\label{fig-ccnnN}
\end{figure*}

Based on Table \ref{ccnnn} and Fig. \ref{fig-ccnnN}, 
we begin by discussing the $cc[nn]\bar{n}$ and $cc[nn]\bar{s}$ subsystems.
For the $cc[nn]\bar{n}$ states with diquark isospin $I_{[nn]}=0$, their total isospin can only couple to $I=1/2$.
For the $cc[nn]\bar{n}$ states with $I_{[nn]}=1$, their total isospin can couple to $I=3/2$ and $I=1/2$.
Here, the $I=3/2$ states and the $I=1/2$ states with $I_{[nn]}=1$ possess identical symmetry constraints on their total wave functions, resulting in their degeneracy in mass spectra, rearrangement decay properties, and RMS radii.
The reason is that the interaction potentials (Eq. (\ref{Eq2})) in the Hamiltonian are independent of the isospin quantum number.
Consistent with similar discussions in Ref.\cite{Zhou:2018bkn}, identical mass spectra are obtained for total isospin $I=1/2$ and $3/2$ in the $ccnn\bar{n}$, $bbnn\bar{n}$, and $bcnn\bar{n}$ subsystems with $I_{nn} = 1$.

Following Fig. \ref{fig-ccnnN}, we easily find that the masses of the $I_{[nn]}=0$ states are generally lower than those of the $I_{[nn]}=1$ states.
The reason is mainly that the mass of the diquark [nn] with $I=1$ is 1.39 GeV, which is significantly higher than that of the diquark [nn] with $I=0$ (1.05 GeV).
Our research findings indicate that states with lower isospin quantum numbers are expected to form more compact multiquark states, thus exhibiting lower masses. 
Furthermore, the experimentally observed $T_{cc}^+(3875)$ is also an isospin scalar state.

In the $I_{[nn]}=0$ states, there are two narrow states, whose total widths are both less than 10 MeV: $P_{c^{2}[nn]\bar{n}}\\(4799,1/2,3/2^{-})$ and $P_{c^{2}[nn]\bar{s}}(4849,0,3/2^{-})$.
While both exhibit a wider decay width than the sub-MeV-scale width of the experimentally observed $T_{cc}^+(3875)$, 
they nevertheless retain the characteristic of a narrow hadronic resonance.
This is because $T_{cc}^+(3875)$ lies slightly below the $D^{*+}D^0$ threshold, making the three-body decay $D^0D^0\pi$ its dominant channel and thus leading to an extremely narrow width. 
In contrast, $P_{c^{2}[nn]\bar{n}}(4799,1/2,3/2^{-})$ and $P_{c^{2}[nn]\bar{s}}(4849,0,3/2^{-})$ lie above the $\Lambda_{c}D^{*}$ and $\Lambda_{c}D_{s}^{*}$ decay thresholds, respectively, thereby accessing a larger decay phase space.
Here, $P_{c^{2}[nn]\bar{n}}(4799,1/2,3/2^{-})$ only decays to $\Lambda_{c}D^{*}$ final states.
Its relatively narrow width yields a well-defined resonance peak, 
and this specific decay channel provides a clear signature for experimental searches.
As a result, this state holds promise for experimental detection. 
Furthermore, we recommend that relevant experimental collaborations check for the signal of $P_{c^{2}[nn]\bar{n}}(4799,1/2,3/2^{-})$ in around 4.8 GeV energy window, where its lineshape is expected to be relatively prominent in the $\Lambda_{c}D^{*}$ mass invariant spectrum.
Meanwhile, $\Lambda_c$ and $D^{*}$ can be readily reconstructed in experimental measurements.
A similar situation also happens in $P_{c^{2}[nn]\bar{s}}(4849,0,3/2^{-})$ which only decays to the $\Lambda_{c}D_{s}^{*}$ final state.
Therefore, we suggest that experiments prioritize searching for potential resonance peaks in the $\Lambda_{c}D_{s}^{*}$ invariant mass spectra within the range of 4.8-4.9 GeV.

Regarding the other two $I_{[nn]}=0$ states in the $cc[nn]\bar{n}$ subsystem: $P_{c^{2}[nn]\bar{n}}(4884,1/2,1/2^{-})$ and $P_{c^{2}[nn]\bar{n}}(4695,1/2,1/2^{-})$,  
they both decay into the $\Lambda_{c}D^{*}$ and $\Lambda_{c}D$ final states, owing to their large phase space.
Their mass gap reaches approximately 200 MeV, which arises from different color configurations and the mixing between different color-spin configurations.
Furthermore, we observe that relative to the $\bar{3}\otimes3$ color configuration, the $6\otimes\bar{6}$ color configuration yields stronger attractive interactions, thereby leading to lower masses for these resonances.
Moreover, their total decay widths are 43 MeV and 17 MeV, respectively.
Additionally, the corresponding relative partial decay width ratios are as follows:
\begin{eqnarray}
\frac{\Gamma[P_{c^{2}[nn]\bar{n}}(4884,1/2,1/2^{-})\to \Lambda_{c}D^{*}]}{\Gamma[P_{c^{2}[nn]\bar{n}}(4884,1/2,1/2^{-})\to\Lambda_{c}D]}=4,
\end{eqnarray}
and 
\begin{eqnarray}
\frac{\Gamma[P_{c^{2}[nn]\bar{n}}(4695,1/2,1/2^{-})\to \Lambda_{c}D^{*}]}{\Gamma[P_{c^{2}[nn]\bar{n}}(4695,1/2,1/2^{-})\to\Lambda_{c}D]}=\frac{5}{2},
\end{eqnarray}
respectively.
For the other two $I_{[nn]}=0$ states in the $cc[nn]\bar{s}$ subsystem, 
similar discussions about the decay behaviors can be conducted by referring to Table \ref{ccnnn} and Fig. \ref{fig-ccnnN}.

In the $I_{[nn]}=1$ states of the $cc[nn]\bar{s}$ subsystem, there are several relatively narrow states.
Among these, $P_{c^{2}[nn]\bar{s}}(5144,1,5/2^{-})$ is the physical state with the narrowest total width.
It has a total width of about 10 MeV and decays predominantly to $\Sigma^{*}_{c}D^{*}_{s}$ final states.
Despite its larger decay phase space,
the Feynman amplitudes $\mathcal{M}(A\to BC)$ (Eq.~(\ref{amp})) associated with its four quark-interchange decay diagrams (Fig. \ref{fig3}) exhibit opposite signs for the $\Sigma^{*}_{c}D^{*}_{s}$ decay channel.
The contributions among them largely cancel out, thereby leading to a suppression of the decay width.
Consequently, $P_{c^{2}[nn]\bar{s}}(5144,1,5/2^{-})$ is expected to exhibit a more prominent lineshape in relevant experimental measurements, thereby enhancing the probability of its experimental discovery. 
Consequently, we propose that relevant experimental collaborations prioritize searching for potential resonance peaks in the $\Sigma^{*}_{c}D^{*}_{s}$ invariant mass spectrum within the $5.1-5.2$ GeV range. 
Moreover, $P_{c^{2}[nn]\bar{s}}(5139,1,3/2^{-})$ and $P_{c^{2}[nn]\bar{s}}(5115,1,3/2^{-})$ are partner states with the same quantum numbers and similar masses.
Their total widths are 63 and 18 MeV, respectively, while the mass gap between them is only 14 MeV.
These two states can be experimentally distinguished by their total widths and partial width ratios, where the latter for each state are given by:
\begin{eqnarray}\label{width3}
\Gamma_{\Sigma^{*}_{c}D^{*}_{s}}:\Gamma_{\Sigma^{*}_{c}D_{s}}:\Gamma_{\Sigma_{c}D^{*}_{s}}=5:2:3, \nonumber\\
\Gamma_{\Sigma^{*}_{c}D^{*}_{s}}:\Gamma_{\Sigma^{*}_{c}D_{s}}:\Gamma_{\Sigma_{c}D^{*}_{s}}=1:3:7,
\end{eqnarray}
respectively.
From the above ratios, we notice that $\Sigma^{*}_{c}D^{*}_{s}$ channel is the dominant decay channel for $P_{c^{2}[nn]\bar{s}}(5139,1,3/2^{-})$.
In contrast, the $\Sigma^{*}_{c}D^{*}_{s}$ channel is suppressed for $P_{c^{2}[nn]\bar{s}}(5115,1,3/2^{-})$, 
which predominantly decays to the $\Sigma_{c}D^{*}_{s}$ final state.

Based on the above research on the typical states, 
and with reference to Table \ref{ccnnn} and 
Fig. \ref{fig-ccnnN}, 
one can perform similar discussions on the decay behaviors of 
other $cc[nn]\bar{n}$ and $cc[nn]\bar{s}$ states, 
and further explore their characteristics in-depth.
\\

\subsection{The $cc[ss]\bar{n}$ and $cc[ss]\bar{s}$ subsystems}
\begin{figure*}[htbp]
\begin{tabular}{c}
\includegraphics[width=460pt]{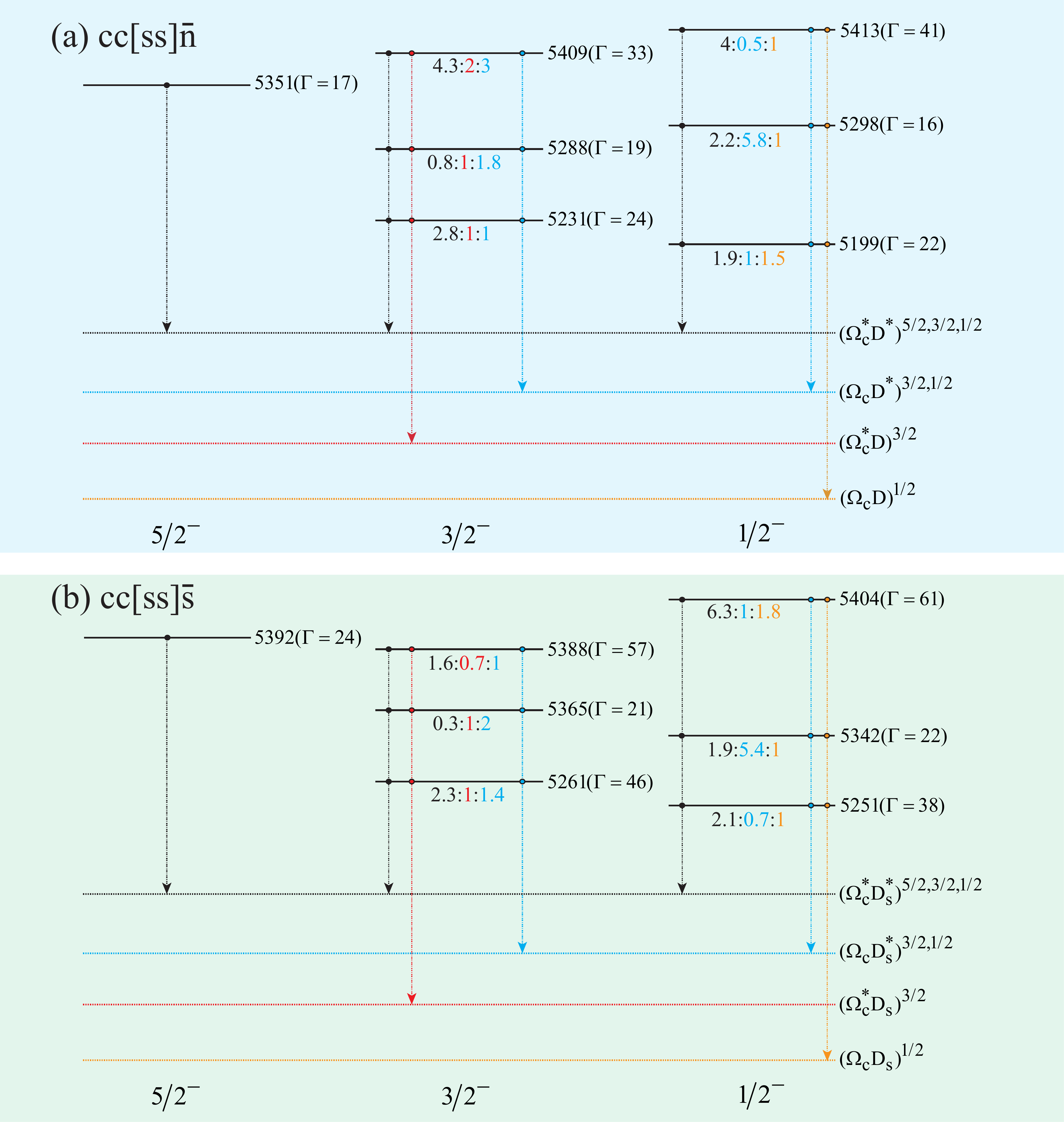}\\
\end{tabular}
\caption{
Relative positions for the 
$cc[ss]\bar{n}$ (a) and $cc[ss]\bar{s}$ (b) pentaquarks labeled with horizontal solid lines,  e.g. $5351~(\Gamma=17)$ represents the mass and total decay width of the corresponding state (units: MeV).
The numbers below the horizontal lines, e.g. $4.3:2:3$, represent the relative branching ratios of the corresponding state.
The dotted lines denote various $S$-wave baryon-meson thresholds, and the superscripts of the labels, e.g. $(\Omega^{*}_{c}D^{*})^{5/2,3/2,1/2}$, represent the possible total angular momenta of the channels.
The solid dots of different colors where the vertical dashed lines with arrows intersect the horizontal solid lines represent the allowed rearranged S-wave decay processes. 
If a vertical dashed line with an arrow intersects a horizontal solid line without a solid dot, it means that corresponding decay process is forbidden for relevant state.
}\label{fig-ccssN}
\end{figure*}

Next, we discuss the $cc[ss]\bar{n}$ and $cc[ss]\bar{s}$ subsystems.
Since they have exactly the same symmetry constraints as the $cc[nn]\bar{n}$ and $cc[nn]\bar{s}$ subsystems with $I_{[nn]}=1$, the number of allowed states is also identical.
According to Table \ref{ccssn} and Fig.\ref{fig-ccssN},
we find that most states are relatively broad states, with their total decay widths exceeding 20 MeV.
Among these, $P_{c^{2}[ss]\bar{n}}(5298,1/2,1/2^{-})$ is the narrowest state with a total width of 16 MeV, and its partial width ratio is given by:
\begin{eqnarray}\label{width4}
\Gamma_{\Omega^{*}_{c}D^{*}}:\Gamma_{\Omega_{c}D^{*}}:\Gamma_{\Omega_{c}D}=2.2:5.8:1.
\end{eqnarray}
We further identify the $\Omega_{c}D^{*}$ decay channel as its primary decay channel.
Consequently, we propose that relevant experimental collaborations searches focus their searches on the signal of $P_{c^{2}[ss]\bar{n}}(5298,1/2,1/2^{-})$ in the $\Omega_{c}D^{*}$ mass invariant spectrum, within an energy window centered at approximately 5.3 GeV.
Another relatively narrow state, $P_{c^{2}[ss]\bar{n}}(5288,1/2,3/2^{-})$, has a total width of 19 MeV, and its partial width ratio is given by:
\begin{eqnarray}\label{width5}
\Gamma_{\Omega^{*}_{c}D^{*}}:\Gamma_{\Omega^{*}_{c}D}:\Gamma_{\Omega_{c}D}=0.8:1:1.8.
\end{eqnarray}
Thus, the $\Omega_{c}D$ decay channel is its dominant decay mode.

The $P_{c^{2}[ss]\bar{s}}(5388,0,3/2^{-})$ and $P_{c^{2}[ss]\bar{s}}(5165,0,3/2^{-})$ states are partner states that have the same quantum numbers and similar masses.
With total widths of 57 and 21 MeV, respectively, their mass gap is approximately 20 MeV; however, this value is relatively small compared to the variation in their decay widths.
Notably, $\Omega^{*}_{c}D_{s}^{*}$ and $\Omega_{c}D_{s}^{*}$ 
decay channels serve as their dominant modes, respectively.
Although theoretically we can distinguish them by their total decay widths, the branching ratios, and dominant decay channels,
current experimental detectors still face great challenges in distinguishing degenerate states with a mass difference ($\Delta M$) of 20 MeV. 
Based on the above research on typical several states, and with reference to the results in Table \ref{ccssn} and Fig.\ref{fig-ccssN}, 
one can perform similar discussions on the decay behaviors of other $cc[ss]\bar{n}$ and $cc[ss]\bar{s}$ states, 
and further investigate their characteristics in-depth.

Moreover, we notice that the physical states predominantly of the 
$|(cc)^{3_{c}}([qq]\bar{q})^{3_{c}}_{1/2}\rangle$ configuration are relatively narrow compared with other states.
Similar phenomena also occur in $cc[nn]\bar{n}$, $cc[nn]\bar{s}$,
$cc[ns]\bar{n}$, and $cc[ns]\bar{s}$ subsystems.
Thus, we suggest that relevant experimental searches prioritize the identification of the physical states predominantly of the 
$|(cc)^{3_{c}}([qq]\bar{q})^{3_{c}}_{1/2}\rangle$ configuration. 
Compared with the $cc[nn]\bar{n}$ and $cc[nn]\bar{s}$ states, the $cc[ss]\bar{n}$ and $cc[ss]\bar{s}$ states exhibit systematically higher masses and broader total decay widths—making their resonance peak lineshapes less pronounced against the inherent experimental background.
Furthermore, in high-energy collision experiments,  the production rate of $s$-quarks is inherently lower than that of $u$- and $d$-quarks. 
Thus, we propose that experimental searches prioritize the narrow-width states in the $cc[nn]\bar{n}$ and $cc[nn]\bar{s}$ subsystems.
\\

\begin{table*}[t]
\centering
\caption{
The numerical results of the mass spectrum, the mass contributions of each Hamiltonian part (in MeV), the root-mean-square radii (in fm), and the partial and total decay widths of the rearrangement decay (in MeV) for the $cc[ss]\bar{n}$ and $cc[ss]\bar{s}$ pentaquarks. 
}\label{ccssn}
\begin{lrbox}{\tablebox}
\renewcommand\arraystretch{2.1}
\renewcommand\tabcolsep{0.75 pt}
\begin{tabular}{ccc|ccc|cccccc|rrrr|r}
\toprule[1.50pt]
\toprule[0.50pt]
\multicolumn{3}{l|}{$cc[ss]\bar{n}$}&
\multicolumn{3}{c|}{Internal contribution}& 
\multicolumn{6}{c|}{RMS Radius}&\multicolumn{5}{c}{Rearrangement decay properties}\\
\Xcline{4-17}{0.3pt}
\multirow{2}*{$I[J^{P}]$}&\multirow{2}*{Configuration}&\multirow{2}*{Mass}&\multirow{2}*{$\langle T \rangle$}
&\multirow{2}*{$\langle V^{\rm Con} \rangle$}
&\multirow{2}*{$\langle V^{\rm SS} \rangle$}
&\multirow{2}*{$R_{12}$}&\multirow{2}*{$R_{34}$}
&$R_{13}$&$R_{23}$
&\multirow{2}*{$R_{12-34}$}
&$R_{13-24}$
&\multirow{2}*{$\Omega^{*}_{c}D^{*}$}
&\multirow{2}*{$\Omega^{*}_{c}D$}
&\multirow{2}*{$\Omega_{c}D^{*}$}
&\multirow{2}*{$\Omega_{c}D$}
&\multicolumn{1}{c}{\multirow{2}*{$\Gamma_{sum}$}}\\
&&&&&&&&$R_{14}$&$R_{24}$&&$R_{14-23}$&&&&&\\
\bottomrule[1.00pt]
\multirow{1}*{$1/2[5/2^{-}]$}&
\multirow{1}*{$
|(cc)^{I=0,\bar{3}_{c}}_{s=1}[(ss)^{I=0,\bar{3}_{c}}_{s=1}\bar{n}]^{I=1/2,3_{c}}_{s=3/2}\rangle^{I=1/2}_{s=5/2}$}
&5350.7&1153.9&-860.7&44.6&1.16&1.49&1.22&1.53&1.35&1.25&16.6&&&&16.6\\
\multirow{3}*{$1/2[3/2^{-}]$}&
\multirow{3}*{$\begin{pmatrix}
|(cc)^{I=0,\bar{3}_{c}}_{s=1}[(ss)^{I=0,\bar{3}_{c}}_{s=1}\bar{n}]^{I=1/2,3_{c}}_{s=3/2}\rangle^{I=1/2}_{s=3/2}
\\
|(cc)^{I=0,\bar{3}_{c}}_{s=1}[(ss)^{I=0,\bar{3}_{c}}_{s=1}\bar{n}]^{I=1/2,3_{c}}_{s=1/2}\rangle^{I=1/2}_{s=3/2}
\\
|(cc)^{I=0,6_{c}}_{s=0}[(ss)^{I=1,\bar{3}_{c}}_{s=1}\bar{n}]^{I=1/2,\bar{6}_{c}}_{s=3/2}\rangle^{I=1/2}_{s=3/2}
\\
\end{pmatrix}$}
&\multirow{3}*{$\begin{pmatrix}
5408.6\\5287.6\\5231.3
\end{pmatrix}$}
&1182.0&-844.8&-3.7&1.35&1.36&1.46&1.50&1.18&1.37&15.5&7.2&10.7&&33.4\\
&&&1176.4&-883.5&3.5&1.16&1.48&1.21&1.53&1.33&1.25&4.1&5.2&9.5&&18.8\\
&&&1198.8&-905.3&-34.3&1.16&1.46&1.21&1.51&1.33&1.24&13.9&5.3&5.0&&24.2\\
\multirow{3}*{$1/2[1/2^{-}]$}&
\multirow{3}*{$\begin{pmatrix}
|(cc)^{I=0,\bar{3}_{c}}_{s=1}[(ss)^{I=0,\bar{3}_{c}}_{s=1}\bar{n}]^{I=1/2,3_{c}}_{s=3/2}\rangle^{I=1/2}_{s=1/2}
\\
|(cc)^{I=0,\bar{3}_{c}}_{s=1}[(nn)^{I=1,\bar{3}_{c}}_{s=1}\bar{n}]^{I=3/2,3_{c}}_{s=1/2}\rangle^{I=1/2}_{s=1/2}
\\
|(cc)^{I=0,6_{c}}_{s=1}[(nn)^{I=1,\bar{3}_{c}}_{s=1}\bar{n}]^{I=3/2,\bar{6}_{c}}_{s=1/2}\rangle^{I=1/2}_{s=1/2}
\\
\end{pmatrix}$}
&\multirow{3}*{$\begin{pmatrix}
5413.3\\5298.0\\5199.3
\end{pmatrix}$}
&1167.8&-830.5&20.0&1.35&1.47&1.18&1.51&1.18&1.37&29.4&&4.0&7.5&40.9\\
&&&1188.7&-895.4&-16.8&1.16&1.47&1.21&1.52&1.34&1.24&4.0&&10.4&1.8&16.2\\
&&&1208.5&-917.7&-54.5&1.15&1.46&1.20&1.51&1.32&1.24&9.4&&4.9&7.5&21.8\\
\bottomrule[1.0pt]
\multicolumn{3}{l|}{$cc[ss]\bar{s}$}&\multirow{2}*{$\langle T \rangle$}
&\multirow{2}*{$\langle V^{\rm Con} \rangle$}
&\multirow{2}*{$\langle V^{\rm SS} \rangle$}
&\multirow{2}*{$R_{12}$}&\multirow{2}*{$R_{34}$}
&$R_{13}$&$R_{14}$
&\multirow{2}*{$R_{12-34}$}
&$R_{13-24}$
&\multirow{2}*{$\Omega^{*}_{c}D_{s}^{*}$}
&\multirow{2}*{$\Omega^{*}_{c}D_{s}$}
&\multirow{2}*{$\Omega_{c}D_{s}^{*}$}
&\multirow{2}*{$\Omega_{c}D_{s}$}
&\multicolumn{1}{c}{\multirow{2}*{$\Gamma_{sum}$}}\\
\multirow{1}*{$I[J^{P}]$}&\multirow{1}*{Configuration}&\multirow{1}*{Mass}&&&&&&$R_{23}$&$R_{24}$&&$R_{14-23}$&&&&&\\
\bottomrule[1.00pt]
\multirow{1}*{$0[5/2^{-}]$}&
\multirow{1}*{$
|(cc)^{I=0,\bar{3}_{c}}_{s=1}[(ss)^{I=0,\bar{3}_{c}}_{s=1}\bar{s}]^{I=0,3_{c}}_{s=3/2}\rangle^{I=0}_{s=5/2}$}
&5391.5&1091.8&-1079.4&39.2&1.16&1.35&1.22&1.39&1.31&1.23&24.4&&&&24.4\\
\multirow{3}*{$0[3/2^{-}]$}&
\multirow{3}*{$\begin{pmatrix}
|(cc)^{I=0,\bar{3}_{c}}_{s=1}[(ss)^{I=0,\bar{3}_{c}}_{s=1}\bar{s}]^{I=0,3_{c}}_{s=3/2}\rangle^{I=0}_{s=3/2}
\\
|(cc)^{I=0,\bar{3}_{c}}_{s=1}[(ss)^{I=0,\bar{3}_{c}}_{s=1}\bar{s}]^{I=0,3_{c}}_{s=1/2}\rangle^{I=0}_{s=3/2}
\\
|(cc)^{I=0,6_{c}}_{s=0}[(ss)^{I=0,\bar{3}_{c}}_{s=1}\bar{s}]^{I=0,\bar{6}_{c}}_{s=3/2}\rangle^{I=0}_{s=3/2}
\\
\end{pmatrix}$}
&\multirow{3}*{$\begin{pmatrix}
5388.2\\5365.4\\5260.7
\end{pmatrix}$}
&1109.6&-1111.8&-2.5&1.33&1.35&1.18&1.36&1.16&1.33&28.1&11.8&17.2&&57.1\\
&&&1112.7&-1100.6&5.0&1.15&1.35&1.21&1.38&1.30&1.22&2.0&6.5&12.7&&21.2\\
&&&1135.8&-1123.1&-30.4&1.15&1.33&1.20&1.37&1.30&1.22&22.7&9.8&13.8&&46.3\\
\multirow{3}*{$0[1/2^{-}]$}&
\multirow{3}*{$\begin{pmatrix}
|(cc)^{I=0,\bar{3}_{c}}_{s=1}[(ss)^{I=0,\bar{3}_{c}}_{s=1}\bar{s}]^{I=0,3_{c}}_{s=3/2}\rangle^{I=0}_{s=1/2}
\\
|(cc)^{I=0,\bar{3}_{c}}_{s=1}[(ss)^{I=0,\bar{3}_{c}}_{s=1}\bar{s}]^{I=0,3_{c}}_{s=1/2}\rangle^{I=0}_{s=1/2}
\\
|(cc)^{I=0,6_{c}}_{s=1}[(ss)^{I=0,\bar{3}_{c}}_{s=1}\bar{s}]^{I=0,\bar{6}_{c}}_{s=1/2}\rangle^{I=0}_{s=1/2}
\\
\end{pmatrix}$}
&\multirow{3}*{$\begin{pmatrix}
5403.8\\5341.6\\5251.2
\end{pmatrix}$}
&1095.7&-1097.8&18.2&1.33&1.36&1.18&1.36&1.16&1.33&42.5&&6.7&12.0&61.1\\
&&&1127.0&-1114.5&-17.8&1.15&1.33&1.21&1.37&1.30&1.22&4.9&&14.0&2.6&21.6\\
&&&1144.8&-1132.1&-46.4&1.15&1.33&1.20&1.36&1.29&1.22&21.2&&6.9&10.1&38.3\\
\bottomrule[0.50pt]
\bottomrule[1.50pt]
\end{tabular}
\end{lrbox}\scalebox{0.81}{\usebox{\tablebox}}
\end{table*}

\subsection{The $cc[ns]\bar{n}$ and $cc[ns]\bar{s}$ subsystems}

\begin{table*}[t]
\centering
\caption{
The numerical results of the mass spectrum, the mass contributions of each Hamiltonian part (in MeV), the root-mean-square radii (in fm), and the partial and total decay widths of the rearrangement decay (in MeV) for the $cc[ns]\bar{n}$ and $cc[ns]\bar{s}$ pentaquarks. 
}\label{ccnsn}
\begin{lrbox}{\tablebox}
\renewcommand\arraystretch{2}
\renewcommand\tabcolsep{0.75 pt}
\begin{tabular}{ccc|ccc|cccccc|rrrrrr|r}
\toprule[1.50pt]
\toprule[0.50pt]
\multicolumn{3}{l|}{$cc[ns]\bar{n}$}&
\multicolumn{3}{c|}{Internal contribution}& 
\multicolumn{6}{c|}{RMS Radius}&\multicolumn{7}{c}{Rearrangement decay properties}\\
\Xcline{4-19}{0.3pt}
\multirow{2}*{$I[J^{P}]$}&\multirow{2}*{Configuration}&\multirow{2}*{Mass}&\multirow{2}*{$\langle T \rangle$}
&\multirow{2}*{$\langle V^{\rm Con} \rangle$}
&\multirow{2}*{$\langle V^{\rm SS} \rangle$}
&\multirow{2}*{$R_{12}$}&\multirow{2}*{$R_{34}$}
&$R_{13}$&$R_{23}$
&\multirow{2}*{$R_{12-34}$}
&$R_{13-24}$
&\multirow{2}*{$\Xi^{*}_{c}D^{*}$}
&\multirow{2}*{$\Xi^{*}_{c}D$}
&\multirow{2}*{$\Xi'_{c}D^{*}$}
&\multirow{2}*{$\Xi'_{c}D$}
&\multirow{2}*{$\Xi_{c}D^{*}$}
&\multirow{2}*{$\Xi_{c}D$}
&\multicolumn{1}{c}{\multirow{2}*{$\Gamma_{sum}$}}\\
&&&&&&&&$R_{14}$&$R_{24}$&&$R_{14-23}$&&&&&&&\\
\bottomrule[1.00pt]
\multirow{1}*{$1(0)[5/2^{-}]$}&
\multirow{1}*{$
|(cc)^{I=0,\bar{3}_{c}}_{s=1}[(ns)^{I=1/2,\bar{3}_{c}}_{s=1}\bar{n}]^{I=1(0),3_{c}}_{s=3/2}\rangle^{I=1(0)}_{s=5/2}$}&
5236.5&1155.8&-847.5&45.6&1.16&1.35&1.23&1.53&1.35&1.26&13.9&&&&&&13.9\\
\multirow{4}*{$1(0)[3/2^{-}]$}&
\multirow{4}*{$\begin{pmatrix}
|(cc)^{I=0,\bar{3}_{c}}_{s=1}[(ns)^{I=1/2,\bar{3}_{c}}_{s=1}\bar{n}]^{I=1(0),3_{c}}_{s=3/2}\rangle^{I=1(0)}_{s=3/2}
\\
|(cc)^{I=0,\bar{3}_{c}}_{s=1}[(ns)^{I=1/2,\bar{3}_{c}}_{s=1}\bar{n}]^{I=1(0),3_{c}}_{s=1/2}\rangle^{I=1(0)}_{s=3/2}
\\
|(cc)^{I=0,6_{c}}_{s=0}[(ns)^{I=1/2,\bar{3}_{c}}_{s=1}\bar{n}]^{I=1(0),\bar{6}_{c}}_{s=3/2}\rangle^{I=1(0)}_{s=3/2}
\\
|(cc)^{I=0,\bar{3}_{c}}_{s=1}[(ns)^{I=1/2,\bar{3}_{c}}_{s=0}\bar{n}]^{I=1(0),3_{c}}_{s=1/2}\rangle^{I=1(0)}_{s=3/2}
\\
\end{pmatrix}$}
&\multirow{4}*{$\begin{pmatrix}
5277.3\\5208.3\\5098.2\\5059.0
\end{pmatrix}$}
&1183.9&-828.7&-4.2&1.35&1.47&1.19&1.50&1.19&1.38&22.9&9.5&14.2&&&&46.6\\
&&&1178.4&-870.4&4.5&1.16&1.49&1.22&1.53&1.34&1.25&2.4&3.4&6.0&&&&11.8\\
&&&1202.4&-893.9&-36.3&1.34&2.15&1.49&2.29&1.80&1.56&12.6&5.3&6.4&&&&24.3\\
&&&1171.1&-838.5&25.9&1.16&1.49&1.24&1.53&1.37&1.26&&&&&11.7&&11.7\\
\multirow{5}*{$1(0)[1/2^{-}]$}&
\multirow{5}*{$\begin{pmatrix}
|(cc)^{I=0,\bar{3}_{c}}_{s=1}[(ns)^{I=1/2,\bar{3}_{c}}_{s=1}\bar{n}]^{I=1(0),3_{c}}_{s=3/2}\rangle^{I=1(0)}_{s=1/2}
\\
|(cc)^{I=0,\bar{3}_{c}}_{s=1}[(ns)^{I=1/2,\bar{3}_{c}}_{s=1}\bar{n}]^{I=1(0),3_{c}}_{s=1/2}\rangle^{I=1(0)}_{s=1/2}
\\
|(cc)^{I=0,6_{c}}_{s=0}[(ns)^{I=1/2,\bar{3}_{c}}_{s=1}\bar{n}]^{I=1(0),\bar{6}_{c}}_{s=1/2}\rangle^{I=1(0)}_{s=1/2}
\\
|(cc)^{I=0,\bar{3}_{c}}_{s=1}[(ns)^{I=1/2,\bar{3}_{c}}_{s=0}\bar{n}]^{I=1(0),3_{c}}_{s=1/2}\rangle^{I=1(0)}_{s=1/2}
\\
|(cc)^{I=0,6_{c}}_{s=0}[(ns)^{I=1/2,\bar{3}_{c}}_{s=0}\bar{n}]^{I=1(0),\bar{6}_{c}}_{s=1/2}\rangle^{I=1(0)}_{s=1/2}
\\
\end{pmatrix}$}
&\multirow{5}*{$\begin{pmatrix}
5301.0\\5180.2\\5137.1\\5084.7\\4945.9
\end{pmatrix}$}
&1168.8&-813.4&-20.9&1.35&1.48&1.19&1.51&1.19&1.38&33.5&&4.2&7.7&&&45.3\\
&&&1192.8&-884.4&-19.6&1.16&1.47&1.22&1.52&1.35&1.25&3.5&&8.9&1.5&&&13.9\\
&&&1181.8&-797.1&4.2&1.36&1.48&1.20&1.50&1.20&1.40&&&&&38.2&9.6&47.9\\
&&&1212.4&-903.8&-55.7&1.16&1.47&1.21&1.51&1.33&1.24&9.9&&4.5&5.8&&&20.2\\
&&&1207.8&-874.8&-39.8&1.16&1.48&1.23&1.52&1.35&1.25&&&&&15.0&6.6&21.5\\
\bottomrule[1.0pt]
\multicolumn{3}{l|}{$cc[ns]\bar{s}$}&\multirow{2}*{$\langle T \rangle$}
&\multirow{2}*{$\langle V^{\rm Con} \rangle$}
&\multirow{2}*{$\langle V^{\rm SS} \rangle$}
&\multirow{2}*{$R_{12}$}&\multirow{2}*{$R_{34}$}
&$R_{13}$&$R_{14}$
&\multirow{2}*{$R_{12-34}$}
&$R_{13-24}$
&\multirow{2}*{$\Xi^{*}_{c}D_{s}^{*}$}
&\multirow{2}*{$\Xi^{*}_{c}D_{s}$}
&\multirow{2}*{$\Xi'_{c}D_{s}^{*}$}
&\multirow{2}*{$\Xi'_{c}D_{s}$}
&\multirow{2}*{$\Xi_{c}D_{s}^{*}$}
&\multirow{2}*{$\Xi_{c}D_{s}$}
&\multicolumn{1}{c}{\multirow{2}*{$\Gamma_{sum}$}}\\
\multirow{1}*{$I[J^{P}]$}&\multirow{1}*{Configuration}&\multirow{1}*{Mass}&&&&&&$R_{23}$&$R_{24}$&&$R_{14-23}$&&&&&&&\\
\bottomrule[1.00pt]
\multirow{1}*{$1/2[5/2^{-}]$}&
\multirow{1}*{$
|(cc)^{I=0,\bar{3}_{c}}_{s=1}[(ns)^{I=1/2,\bar{3}_{c}}_{s=1}\bar{s}]^{I=1/2,3_{c}}_{s=3/2}\rangle^{I=1/2}_{s=5/2}$}
&5278.3&1094.2&-1065.6&40.1&1.16&1.36&1.22&1.38&1.32&1.23&19.6&&&&&&19.6\\
\multirow{4}*{$1/2[3/2^{-}]$}&
\multirow{4}*{$\begin{pmatrix}
|(cc)^{I=0,\bar{3}_{c}}_{s=1}[(ns)^{I=1/2,\bar{3}_{c}}_{s=1}\bar{s}]^{I=1/2,3_{c}}_{s=3/2}\rangle^{I=1/2}_{s=3/2}
\\
|(cc)^{I=0,\bar{3}_{c}}_{s=1}[(ns)^{I=1/2,\bar{3}_{c}}_{s=1}\bar{s}]^{I=1/2,3_{c}}_{s=1/2}\rangle^{I=1/2}_{s=3/2}
\\
|(cc)^{I=0,\bar{3}_{c}}_{s=1}[(ns)^{I=1/2,\bar{3}_{c}}_{s=0}\bar{s}]^{I=1/2,3_{c}}_{s=1/2}\rangle^{I=1/2}_{s=3/2}
\\
|(cc)^{I=0,6_{c}}_{s=0}[(ns)^{I=1/2,\bar{3}_{c}}_{s=1}\bar{s}]^{I=1/2,\bar{6}_{c}}_{s=3/2}\rangle^{I=1/2}_{s=3/2}
\\
\end{pmatrix}$}
&\multirow{4}*{$\begin{pmatrix}
5274.3\\5251.1\\5147.8\\5103.9
\end{pmatrix}$}
&1111.4&-1095.4&-2.9&1.33&1.23&1.36&1.36&1.16&1.33&27.4&11.2&16.8&&&&55.4\\
&&&1115.2&-1086.9&5.7&1.15&1.36&1.22&1.37&1.31&1.23&1.6&4.8&10.2&&&&16.6\\
&&&1139.5&-1110.6&-31.7&1.15&1.33&1.22&1.37&1.31&1.22&21.4&8.9&12.8&&&&43.1\\
&&&1110.2&-1055.4&21.7&1.16&1.36&1.24&1.38&1.33&1.23&&&&&10.8&&10.8\\
\multirow{5}*{$1/2[1/2^{-}]$}&
\multirow{5}*{$\begin{pmatrix}
|(cc)^{I=0,\bar{3}_{c}}_{s=1}[(ns)^{I=1/2,\bar{3}_{c}}_{s=1}\bar{s}]^{I=1(0),3_{c}}_{s=3/2}\rangle^{I=1/2}_{s=1/2}
\\
|(cc)^{I=0,\bar{3}_{c}}_{s=1}[(ns)^{I=1/2,\bar{3}_{c}}_{s=1}\bar{s}]^{I=1/2,3_{c}}_{s=1/2}\rangle^{I=1/2}_{s=1/2}
\\
|(cc)^{I=0,\bar{3}_{c}}_{s=1}[(ns)^{I=1/2,\bar{3}_{c}}_{s=0}\bar{s}]^{I=1/2,3_{c}}_{s=1/2}\rangle^{I=1/2}_{s=1/2}
\\
|(cc)^{I=0,6_{c}}_{s=0}[(ns)^{I=1/2,\bar{3}_{c}}_{s=1}\bar{s}]^{I=1/2,\bar{6}_{c}}_{s=1/2}\rangle^{I=1/2}_{s=1/2}
\\
|(cc)^{I=0,6_{c}}_{s=0}[(ns)^{I=1/2,\bar{3}_{c}}_{s=0}\bar{s}]^{I=1/2,\bar{6}_{c}}_{s=1/2}\rangle^{I=1/2}_{s=1/2}
\\
\end{pmatrix}$}
&\multirow{5}*{$\begin{pmatrix}
5291.4\\5225.3\\5137.4\\5132.6\\4997.0
\end{pmatrix}$}
&1096.8&-1080.7&18.8&1.33&1.36&1.19&1.36&1.17&1.34&45.1&&6.2&11.3&&&62.6\\
&&&1131.2&-1102.5&-20.0&1.15&1.34&1.22&1.37&1.31&1.22&3.6&&10.4&1.8&&&15.8\\
&&&1148.3&-1119.5&-47.4&1.15&1.34&1.21&1.36&1.30&1.22&21.0&&8.1&11.3&&&40.4\\
&&&1109.4&-1063.2&4.4&1.34&1.37&1.21&1.31&1.17&1.35&&&&&47.5&12.3&59.8\\
&&&1143.7&-1088.7&-31.0&1.15&1.36&1.23&1.37&1.31&1.23&&&&&33.0&9.7&42.7\\
\bottomrule[0.50pt]
\bottomrule[1.50pt]
\end{tabular}
\end{lrbox}\scalebox{0.78}{\usebox{\tablebox}}
\end{table*}

\begin{figure*}[htbp]
\begin{tabular}{c}
\includegraphics[width=455pt]{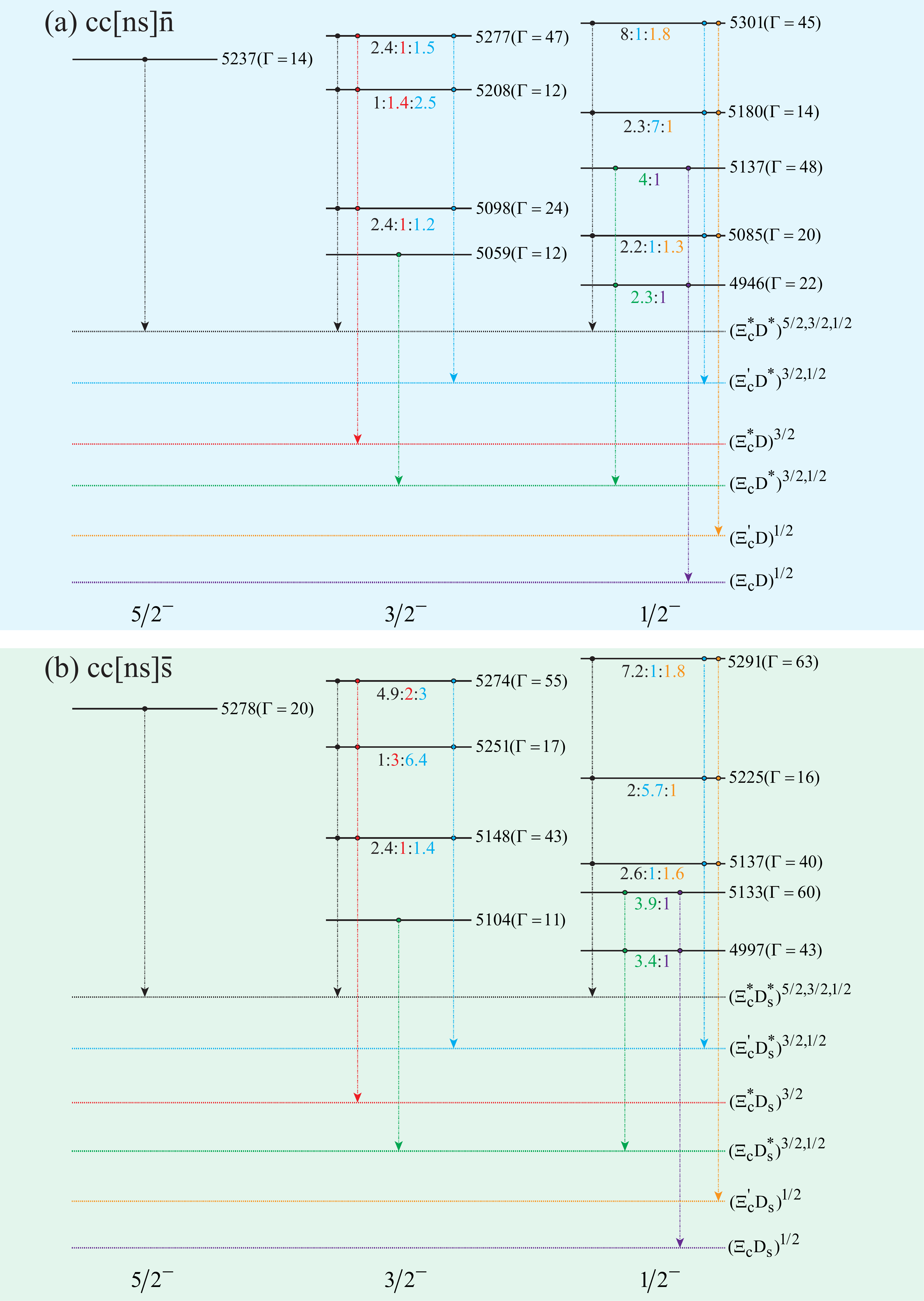}\\
\end{tabular}
\caption{
Relative positions for the $cc[ns]\bar{n}$ (a) and $cc[ns]\bar{s}$ (b) pentaquarks labeled with horizontal solid lines, the labels, e.g. $5237(\Gamma=14)$ represents the mass and total decay width of the corresponding state (units: MeV).
The numbers below the horizontal lines, the labels, e.g. $2.4:1:1.5$, represent the relative branching ratios of the corresponding state.
The dotted lines denote various $S$-wave baryon-meson thresholds, and the superscripts of the labels, e.g. $(\Xi^{*}_{c}D^{*})^{5/2,3/2,1/2}$, represent the possible total angular momenta of the channels.
The solid dots of different colors where the vertical dashed lines with arrows intersect the horizontal solid lines represent the allowed rearranged S-wave decay processes. 
If a vertical dashed line with an arrow intersects a horizontal solid line without a solid dot, it means that corresponding decay process is forbidden for relevant state.
}\label{fig-ccnsN}
\end{figure*}

Finally, we focus on the $cc[ns]\bar{n}$ and $cc[ns]\bar{s}$ subsystems.
For the $cc[ns]\bar{n}$ states with diquark isospin $I_{[ns]}=1/2$, their total isospin can be coupled to $I=1$ and $I=0$.
Similar to the $I=3/2$ and $I=1/2$ states (with $I_{[nn]}=1$) in the $cc[nn]\bar{n}$ subsystem, 
the $I=1$ and $I=0$ states of the $cc[ns]\bar{n}$ subsystem also possess identical symmetry constraints on their total wave functions, resulting in their degeneracy in mass spectra, rearrangement decay properties, and RMS radii.

Referring to Table \ref{ccnsn} and Fig. \ref{fig-ccnsN}, 
it is evident that the masses of states containing the $(ns)^{I=1/2}_{s=0}$ diquark are generally lower than those of states containing the $(ns)^{I=1/2}_{s=1}$ diquark.
The reason is mainly that the mass of $(ns)^{I=1/2}_{s=1}$ diquark is 1.54 GeV, which is significantly higher than that of the $(ns)^{I=1/2}_{s=0}$ diquark (1.36 GeV).

In the $cc[ns]\bar{n}$ and $cc[ns]\bar{s}$ subsystems,
there are three relatively narrow states: $P_{c^{2}[ns]\bar{n}}(5208,1(0),3/2^{-})$, $P_{c^{2}[ns]\bar{n}}(5059,1(0),3/2^{-})$, and 
$P_{c^{2}[ns]\bar{s}}(5104,1/2,3/2^{-})$.
Meanwhile, we note that these three physical states all exhibit a predominant $|(cc)^{3_{c}}([qq]\bar{q})^{3_{c}}_{1/2}\rangle$ configuration.
Among these, $P_{c^{2}[ns]\bar{n}}(5059,1(0),3/2^{-})$ and 
$P_{c^{2}[ns]\bar{s}}(5104,1/2,3/2^{-})$ exhibit a total decay width of around 11 MeV each and only decay to the $\Xi_{c}D^{*}$ and $\Xi_{c}D_{s}^{*}$ final states, respectively.
Consequently, we propose that experimental searches prioritize the identification of potential resonance peaks in the $\Xi_{c}D^{*}$ and $\Xi_{c}D_{s}^{*}$ invariant mass spectra within the $5000-5100$ MeV and $5050-5150$ MeV ranges, respectively. 
As for $P_{c^{2}[ns]\bar{n}}(5208,1(0),3/2^{-})$, 
it has a total width of 12 MeV, and its partial width ratios are given by:
\begin{eqnarray}\label{width6}
\Gamma_{\Xi^{*}_{c}D^{*}}:\Gamma_{\Xi^{*}_{c}D}:\Gamma_{\Xi'_{c}D^{*}}=1:1.4:2.5.
\end{eqnarray}
Thus, the $\Xi'_{c}D^{*}$ decay channel is its dominant decay mode.
Furthermore, 
we suggest that experiments prioritize the search for potential resonance peaks in the $5150-5250$ MeV range of the $\Xi'_{c}D^{*}$ invariant mass spectrum. 

Moreover, the $P_{c^{2}[ns]\bar{s}}(5137,1/2,1/2^{-})$ and 
$P_{c^{2}[ns]\bar{s}}\\(5133,1/2,1/2^{-})$ states are partner states, which have identical quantum numbers and similar masses.
Their total decay widths are 40 MeV and 60 MeV, respectively, with a mass gap of approximately 4 MeV; thus, their mass gap is negligible compared to their total decay widths.
However, their decay channels are completely different.
Their partial width ratios are expressed as follows:
\begin{eqnarray}\label{width7}
\Gamma_{\Xi^{*}_{c}D_{s}^{*}}:\Gamma_{\Xi'_{c}D_{s}^{*}}:\Gamma_{\Xi'_{c}D_{s}}&=&2.6:1:1.6, \nonumber\\
\Gamma_{\Xi_{c}D_{s}^{*}}:\Gamma_{\Xi D_{s}}&=&3.9:1,
\end{eqnarray}
respectively.
Although they have similar masses and identical quantum numbers, they can be easily distinguished by their decay final states.
The $P_{c^{2}[ns]\bar{s}}(5274,\\1/2,3/2^{-})$ and 
$P_{c^{2}[ns]\bar{s}}(5251,1/2,3/2^{-})$ states are also partner states.
They both have identical quantum numbers and similar masses; however, their total decay widths exhibit significant differences.
Therefore, we can distinguish these two physical states based on the significance of their resonance peak lineshapes.
Based on the above studies on the typical states and with reference to the results presented in Table \ref{ccnsn} and Fig.\ref{fig-ccnsN}, similar discussions on the decay behaviors of other states have been derived.
\\

\section{summary}\label{sec4}
With the experimental discovery and subsequent theoretical investigations of the doubly-charmed baryon $\Xi^{++}_{cc}(3621)$ and the doubly-charmed tetraquark candidate state $T^{+}_{cc}(3875)$, we focus our attention on the doubly-charmed pentaquark system.
Based on the strong predictive power and broad applicability of the diquark model, we extend its application to the doubly-charmed pentaquark system, reducing it to a relatively simple four-body problem.
We investigate this system within the 
heavy quark-heavy quark-diquark-antiquark configuration, aiming to provide more realistic mass spectra for the doubly-charmed pentaquark system.

In this work, we first construct total wave functions including the flavor, spatial, color, and spin parts. 
Next, within the framework of the constituent quark model, 
we systematically calculate the mass spectra of the 
doubly-charmed pentaquark system by employing the Gaussian expansion method.
Finally, we further calculate the corresponding internal mass contributions, root-mean-square (RMS) radii, and rearrangement decay properties.

After considering the mixing between different color-spin configurations—an effect driven by the non-zero off-diagonal elements of the hyperfine interaction potential matrix $\langle V^{\rm SS} \rangle$, which plays a crucial role in mixing configurations of the form $|(cc)^{\rm color_{1}}_{\rm spin_{1}}[(q_{1}q_{2})^{\bar{3}_{c}}_{\rm spin_{2}}\bar{q}_{\frac{1}{2}}]^{\rm color_{2}}_{\rm spin_{3}}\rangle_{\rm spin_{4}}$—we obtain the physical states of the doubly-charmed pentaquark system. 
It is worth emphasizing that the color-spin configuration mixing induces notable mass shifts and larger mass gaps for the physical states compared with their pre-mixing counterparts.
Meanwhile, our theoretical predictions indicate that these physical states have masses in the range of 4.7–5.4 GeV, as detailed in Tables \ref{ccnnn}–\ref{ccnsn}. 
Analysis of internal mass contributions reveals that the kinetic energy $\langle T \rangle$ and confinement potential $\langle V^{\rm Con} \rangle$ are of the same order of magnitude. 
In addition, there exist no stable states in this system: all physical states can undergo rearrangement decay to form a singly-charmed meson and a singly-charmed baryon as the final state. 

Besides the mass spectra, 
we also present the corresponding rms radii,
with most results falling within the range of 1.1–1.6 fm and exhibiting roughly the same order of magnitude.
Meanwhile, our calculations yield the following relations: $\langle r^{2}_{13} \rangle^{\frac{1}{2}}=\langle r^{2}_{23} \rangle^{\frac{1}{2}}$, $\langle r^{2}_{14} \rangle^{\frac{1}{2}}=\langle r^{2}_{24} \rangle^{\frac{1}{2}}$, and $\langle r^{2}_{13-24} \rangle^{\frac{1}{2}}=\langle r^{2}_{14-23} \rangle^{\frac{1}{2}}$, which are in full agreement with our symmetry analysis.
Moreover, our results imply that the spatial distribution between quarks is relatively compact and the internal interactions within the system are strong,
which align with the expectations for the compact pentaquark configuration.

Subsequently, we conducted an in-depth analysis of the rearrangement decay properties. 
Our results show that
most physical states possess total decay widths ranging from 15 to 70 MeV.
Of course, there are still several narrow physical states, some of which have a total decay width of even less than 10 MeV.
The reason is that, despite their larger decay phase space, 
the signs of the Feynman amplitudes $\mathcal{M}(A\to BC)$ from the four quark-interchange diagrams are different. 
This causes their contributions to cancel each other out to a great extent, ultimately suppressing the decay width and leading to the emergence of narrow states.
Moreover, we observe that the physical states predominantly of the 
$|(cc)^{3_{c}}([qq]\bar{q})^{3_{c}}_{1/2}\rangle$ configuration are relatively narrow compared to other states.

Among them, the total decay widths of $P_{c^{2}[nn]\bar{n}}(4799,\\1/2,3/2^{-})$ and $P_{c^{2}[nn]\bar{s}}(4849,0,3/2^{-})$ are 7 MeV and 6 MeV, respectively.
Meanwhile, they only decay to the $\Lambda_{c}D^{*}$ and $\Lambda_{c}D_{s}^{*}$ final states, respectively.
Their relatively narrow width yields a prominent resonance peak,
and this specific decay channel provides a clear signature for experimental searches.
Therefore, we suggest that experimental collaborations should prioritize searches for these two potential resonance peaks, whose lineshapes should be relatively prominent in the $\Lambda_{c}D^{*}$ and $\Lambda_{c}D_{s}^{*}$ invariant mass spectra, respectively.
Relative to $cc[nn]\bar{n}$ and $cc[nn]\bar{s}$ states, 
other states have higher masses and broader total decay widths—rendering their resonance peak lineshapes less distinct against experimental backgrounds. 
Additionally, $s$-quark production rates are experimentally lower than those of $u$- and $d$-quarks. 
Hence, we recommend experimental searches prioritize narrow states in the $cc[nn]\bar{n}$ and $cc[nn]\bar{s}$ subsystems.

In summary, 
our results comprehensively reveal the mass spectra, internal structures, and decay characteristics of doubly-charmed pentaquarks, which can provide some preliminary hints for theorists and experimentalists.
More detailed dynamical studies of these pentaquarks are still imperative. 
Finally, we hope that they will provide valuable perspectives for further theoretical research.
Furthermore, we also hope that this study can inspire the LHCb, BelleII, BESIII and other relevant experimental collaborations to search for doubly-charmed pentaquarks.
Conducting more experimental measurements can not only test our calculated results, but also deepen the understanding of the interactions within pentaquarks.
\\

\section*{Acknowledgements}
This work is supported by the National Nature Science Foundation of China under Grant No.12447172 and 12447155, by the Postdoctoral Fellowship Program of CPSF under Grant No.GZC20240877, 2025M773368, and GZC20240056 and by Shuimu Tsinghua Scholar Program of Tsinghua University under Grant No.2024SM119.

\bibliographystyle{UserDefined}
\bibliography{References}

\end{document}